\documentclass[aps,pre,reprint,showpacs,groupedaddress]{revtex4-1}

\usepackage{tikz}
\usepackage{amsmath}
\usepackage{amssymb}
\usepackage{hyperref}
\usepackage[capitalize]{cleveref}
\hypersetup{
    colorlinks,
    citecolor=blue,
    linkcolor=blue,     urlcolor=blue
}

\newcommand{\Km}{\vec{K}_{\vec{m}}}
\newcommand{\km}{\vec{k}_{\vec{m}}}
\renewcommand{\vec}[1]{\boldsymbol{#1}}
\newcommand{\vgal}{\vec{v}_{gal}}
\newcommand{\nab}{\vec{\nabla'}}
\newcommand{\Dt}[1]{ \frac{\partial #1}{\partial t}}
\newcommand{\mc}[1]{\hat{\mathcal{#1}}}
\newcommand{\xj}{\vec{x}'_{\vec{j}}}
\newcommand{\Xll}{\vec{X}_{\vec{\ell}}}
\newcommand{\Integ}[1]{\int_{-\infty}^{\infty} \!\!\!\!\!\!
  \mathrm{d}#1}
\newcommand{\RInteg}[1]{\int_{0}^{\infty} \!\! \frac{#1\mathrm{d}#1}{(2\pi)^2}}

\begin{document}


\title{Elimination of Numerical Cherenkov Instability in
  flowing-plasma Particle-In-Cell simulations by using Galilean coordinates}

\author{Remi Lehe$^a$}
\email[]{rlehe@lbl.gov}
\author{Manuel Kirchen$^b$}
\author{Brendan B. Godfrey$^{a,c}$}
\author{Andreas R. Maier$^{b}$}
\author{Jean-Luc Vay$^a$}
\affiliation{$^a$ Lawrence Berkeley National Laboratory, Berkeley, CA
  94720, USA \\
$^b$ Center for Free-Electron Laser Science \& Department of Physics,
University of Hamburg, 22761 Hamburg, Germany\\
$^c$ University of Maryland, College Park, MD 20742, USA}


\date{\today}

\begin{abstract}
Particle-In-Cell (PIC) simulations of relativistic flowing plasmas are 
of key interest to several fields of physics (including e.g. laser-wakefield
acceleration, when viewed in a Lorentz-boosted frame), 
but remain sometimes infeasible due to the
well-known numerical Cherenkov instability (NCI). In this article, we
show that, for a plasma drifting at a uniform relativistic velocity, the NCI can be eliminated by simply integrating the PIC
equations in \emph{Galilean coordinates} that follow the plasma 
(also sometimes known as
\emph{comoving coordinates}) within a spectral
analytical framework. 
The elimination of the NCI is verified
empirically and confirmed by a theoretical analysis of the
instability. Moreover, it is shown that this method is applicable both to
Cartesian geometry and to cylindrical geometry with azimuthal Fourier
decomposition. 
\end{abstract}


\maketitle

\section*{Introduction}

Simulating relativistic flowing plasmas is of importance in several
fields of physics, including relativistic astrophysics (e.g. \cite{SpitkovskyApJ2008,KeshetApJ2009}) and
laser-plasma acceleration \cite{TajimaPRL1979}. More precisely,
although in laser-plasma acceleration the plasma is typically at
rest in the laboratory frame, it was shown \cite{VayPRL2007} that simulating the
interaction in a Lorentz-boosted frame -- where the plasma is flowing
with relativistic speed -- reduces computational demands
by orders of magnitude. 

However, despite the interest surrounding simulations of relativistic flowing
plasmas, performing these simulations with Particle-In-Cell (PIC)
algorithms \cite{Hockney1988,Birdsall2004} remains a challenge. This is because a violent numerical instability, known as the
numerical Cherenkov instability (NCI) \cite{MartinsCPC10,VayAAC10,VayJCP11,VayPOPL11,GodfreyJCP1974,GodfreyJCP1975,GodfreyJCP2013,XuCPC2013}, quickly
develops for relativistic plasmas and disrupts the simulation. 

Several solutions have been proposed to mitigate the NCI \cite{GodfreyJCP2014,GodfreyIEEE2014,GodfreyJCP2014b,Godfreyarxiv2014,GodfreyCPC2015,YuCPC2015,YuCPC2015-Circ,Yu-arxiv2016}. Although
these solutions efficiently reduce the numerical instability,
they typically introduce either strong smoothing of the currents and
fields, or arbitrary numerical corrections, which are
tuned specifically against the NCI and go beyond the
natural discretization of the underlying physical equation. Therefore,
it is sometimes unclear to what extent these added corrections could impact the
physics at stake.

For instance, NCI-specific corrections include periodically smoothing 
the electromagnetic field components \cite{MartinsCPC10}, 
using a special time step \cite{VayAAC10,VayJCP11} or
applying a wide-band smoothing of the current components \cite{
  VayAAC10,VayJCP11,VayPOPL11}. Another set of mitigation methods
involve scaling the deposited
currents by a carefully-designed wavenumber-dependent factor
\cite{GodfreyJCP2014,GodfreyIEEE2014} or slightly modifying the
ratio of electric and magnetic fields ($E/B$) before gathering their
value onto the macroparticles 
\cite{GodfreyJCP2014b,Godfreyarxiv2014,GodfreyCPC2015}. 
Yet another set of NCI-specific corrections
\cite{YuCPC2015,YuCPC2015-Circ,Yu-arxiv2016} consists 
in combining a small timestep $\Delta t$, a sharp low-pass spatial filter,
and a spectral or high-order scheme that is tuned so as to
create a small, artificial ``bump'' in the dispersion relation
\cite{YuCPC2015}. While most mitigation methods have only been applied
to Cartesian geometry, this last
set of methods (\cite{YuCPC2015,YuCPC2015-Circ,Yu-arxiv2016}) 
has the remarkable property that it can be applied
\cite{YuCPC2015-Circ} to both Cartesian geometry and
quasi-cylindrical geometry (i.e. cylindrical geometry with
azimuthal Fourier decomposition \cite{Lifschitz,Davidson}). However,
the use of a small timestep proportionally slows down the progress of
the simulation, and the artificial ``bump'' is again an arbitrary correction
that departs from the underlying physics.

By contrast, in \cite{Kirchen2016}, we propose that the NCI can be
eliminated -- with no arbitrary correction -- by simply integrating
the PIC equations in \emph{Galilean coordinates} (also known as
\emph{comoving coordinates}). More precisely, in our
method, the Maxwell equations \emph{in Galilean coordinates} are integrated
analytically, using only natural hypotheses, within the PSATD
framework (Pseudo-Spectral-Analytical-Time-Domain \cite{Haber,VayJCP2013}). 
In the present article, we present the mathematical
derivation and implementation of this \emph{Galilean PSATD} scheme. 
Moreover, we conduct a detailed empirical and theoretical stability
analysis for a \emph{uniform} flowing plasma. On the other hand, 
the practical application of this algorithm to realistic, non-uniform
plasmas (such as e.g. in laser-wakefield
acceleration) is presented in \cite{Kirchen2016}. Overall, our method intrinsically supresses the NCI, does not require a small
timestep, and applies to both Cartesian and quasi-cylindrical geometry.


The outline of the present article is the following. We give an
intuitive explanation of the Galilean scheme in \cref{sec:intuitive},
and then detail its exact implementation for Cartesian geometry in
\cref{sec:cartesian}. We also show \emph{empirically}, in
\cref{sec:cartesian}, that, for a plasma drifting at a uniform 
relativistic velocity, the Galilean scheme suppresses the NCI.
This fact is then confirmed and explained by
a theoretical stability analysis in \cref{sec:stability-analysis}
(again, for Cartesian geometry). Finally, \cref{sec:cylindrical} shows
that the Galilean scheme can also be applied to quasi-cylindrical
geometry, and that it is then equally effective at suppressing the NCI.

\section{\label{sec:intuitive}An intuitive explanation of the Galilean scheme}

\begin{figure}
\begin{tikzpicture}

\draw (-0.5,2) node{\textbf{Standard scheme}};

\begin{scope}[shift={(-2,0)}]
  \fill[fill=blue!30!white] (0.2,0.1) rectangle (2.4,1.4);
  \draw (1.5,0.55) node[anchor=south,blue,text width=2cm]{\textbf{Relativistic plasma}}; 
  \draw[draw=blue,->,thick] (2.4,0.7) -- +(0.4,0) node[blue,above]{$\vec{v}_0$};
  \draw[step=0.5,black,very thin] (0,0) grid (3,1.5);
\end{scope}

\begin{scope}[shift={(-2,-2)}]
  \fill[fill=blue!30!white] (0.4,0.1) rectangle (2.6,1.4);
  \draw[draw=blue,->,thick] (2.6,0.7) -- +(0.4,0);
  \draw[step=0.5,black,very thin] (0,0) grid (3,1.5);
\end{scope}

\begin{scope}[shift={(-2,-4)}]
  \fill[fill=blue!30!white] (0.6,0.1) rectangle (2.8,1.4);
  \draw[draw=blue,->,thick] (2.8,0.7) -- +(0.4,0);
  \draw[step=0.5,black,very thin] (0,0) grid (3,1.5);
\end{scope}

\draw (3.5,2) node{\textbf{Galilean scheme}};

\begin{scope}[shift={(2,0)}]
  \fill[fill=blue!30!white] (0.2,0.1) rectangle (2.4,1.4);
  \draw[draw=blue,->,thick] (2.4,0.7) -- +(0.4,0) node[blue,above]{$\vec{v}_0$};
  \draw[draw=gray,->,thick] (3,0.7) -- +(0.4,0) node[above]{$\vgal$};
  \draw[step=0.5,black,very thin] (0,0) grid (3,1.5);
\end{scope}

\begin{scope}[shift={(2.2,-2)}]
  \fill[fill=blue!30!white] (0.2,0.1) rectangle (2.4,1.4);
  \draw[draw=blue,->,thick] (2.4,0.7) -- +(0.4,0);
  \draw[draw=gray,->,thick] (3,0.7) -- +(0.4,0);
  \draw[step=0.5,black,very thin] (0,0) grid (3,1.5);
\end{scope}

\begin{scope}[shift={(2.4,-4)}]
  \fill[fill=blue!30!white] (0.2,0.1) rectangle (2.4,1.4);
  \draw[draw=blue,->,thick] (2.4,0.7) -- +(0.4,0);
  \draw[draw=gray,->,thick] (3,0.7) -- +(0.4,0);
  \draw[step=0.5,black,very thin] (0,0) grid (3,1.5);
\end{scope}

\draw[->,very thick] (1.5,2) -- (1.5,-4.5) node[below]{Time};

\draw[->] (-2.4,-4.4) -- +(0.5,0) node[below]{$z$};
\draw[->] (-2.4,-4.4) -- +(0,0.5) node[left]{$x$};

\end{tikzpicture}
\caption{\label{fig:schematic}Schematic representation of the Galilean
  scheme. As explained in the text, the Galilean scheme is \emph{not} equivalent to a
moving window.}
\end{figure}
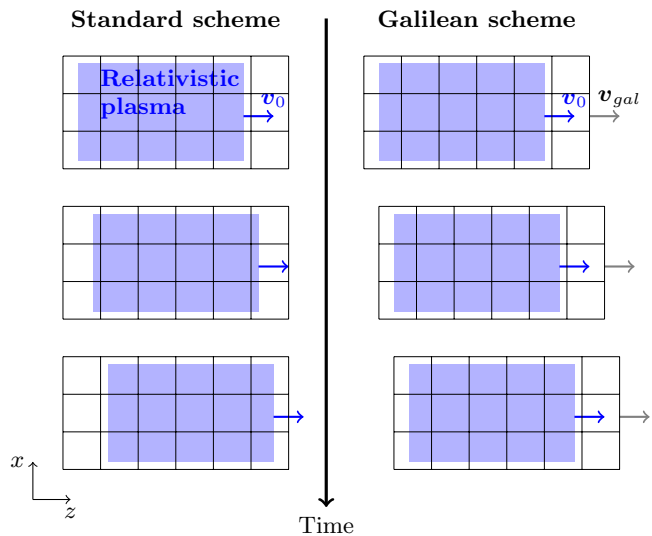

The idea of the proposed scheme is to perform a Galilean change of
coordinates, and to carry out the simulation in the new coordinates:
\begin{equation} 
\label{eq:change-var}
\vec{x}' = \vec{x} - \vgal t 
\end{equation}
where $\vec{x} = x\,\vec{u}_x + y\,\vec{u}_y + z\,\vec{u}_z$ and
$\vec{x}' = x'\,\vec{u}_x + y'\,\vec{u}_y + z'\,\vec{u}_z$ are the
position vectors in the standard and Galilean coordinates
respectively.

We typically choose $\vgal= \vec{v}_0$, where
$\vec{v}_0$ is the speed of the bulk of the relativistic
plasma. In this case, in the Galilean coordinates $\vec{x}'$, the plasma does
not move with respect to the grid -- or, equivalently, in the standard
coordinates $\vec{x}$, the grid moves along with the plasma (as represented
in \cref{fig:schematic}). The heuristic intuition behind this scheme
is that these coordinates should prevent the discrepancy between the Lagrangian and
Eulerian points of view, which gives rise to the NCI \cite{GodfreyJCP1975}.

While in the standard coordinates $\vec{x}$, the equations of particle
motion and the Maxwell
equations have the familiar form
\begin{subequations}
\begin{align}
\frac{d\vec{x}}{dt} &= \frac{\vec{p}}{\gamma m} \label{eq:motion1-standard} \\ 
\frac{d\vec{p}}{dt} &= q \left( \vec{E} +
\frac{\vec{p}}{\gamma m} \times \vec{B} \right) \label{eq:motion2-standard}\\
\frac{\partial \vec{B}}{\partial t} &= -\nabla\times\vec{E} \label{eq:maxwell1-standard}\\
 \frac{1}{c^2}\frac{\partial \vec{E}}{\partial t} &= \nabla\times\vec{B} - \mu_0\vec{j} \label{eq:maxwell2-standard}
\end{align}
\end{subequations}
in the Galilean coordinates $\vec{x}'$, these equations become
\begin{subequations}
\begin{align}
\frac{d\vec{x}'}{dt} &= \frac{\vec{p}}{\gamma m} - \vgal \label{eq:motion1} \\ 
\frac{d\vec{p}}{dt} &= q \left( \vec{E} +
\frac{\vec{p}}{\gamma m} \times \vec{B} \right) \label{eq:motion2}\\
\left( \Dt{\;} - \vgal\cdot\nab\right)\vec{B} &= -\nab\times\vec{E} \label{eq:maxwell1}\\
\frac{1}{c^2}\left( \Dt{\;} - \vgal\cdot\nab\right)\vec{E} &= \nab\times\vec{B} - \mu_0\vec{j} \label{eq:maxwell2}
\end{align}
\end{subequations}
where $\nab$ denotes a spatial derivative with respect to the
Galilean coordinates $\vec{x}'$. The idea of the Galilean scheme is to
design a PIC code which integrates the equations
\cref{eq:motion1,eq:motion2,eq:maxwell1,eq:maxwell2} instead of
\cref{eq:motion1-standard,eq:motion2-standard,eq:maxwell1-standard,eq:maxwell2-standard}. Of
course, physically, these two sets of equations are equivalent, as they are simply connected by a change of
variables. However, we show in this paper that, numerically, these
sets of equations have different stability properties, 
when integrated with the PSATD scheme. 
Indeed, as shown in the next section, one of the unique feature of the
PSATD scheme is that it takes into account the assumed time evolution of the 
current $\vec{j}$ \emph{within one timestep}. This allows us to push
the idea of the coordinate change \cref{eq:change-var} further, by
embedding it into the assumed time evolution of $\vec{j}$ (see
\cref{eq:assumption-standard,eq:assumption} in the next section). As
we will show in \cref{sec:stability-analysis}, this turns out to be
key for the elimination of the NCI.

Before going further, let us remark that the Galilean change of
coordinates \cref{eq:change-var} is a simple translation. Thus, when used in
the context of Lorentz-boosted simulations \cite{VayPRL2007}, it does
of course preserve the relativistic dilatation of space and time which gives rise to the
characteristic computational speedup of the boosted-frame technique.

Another important remark is that the Galilean scheme is \emph{not}
equivalent to a moving window (and in fact the Galilean scheme can be
independently \emph{combined} with a moving window). Whereas in a
moving window, gridpoints are added and removed so as to effectively
translate the boundaries, in the Galilean scheme the gridpoints
\emph{themselves} are translated (and, again, in this case the physical equations
are modified accordingly). In addition, the assumed time evolution of
$\vec{j}$ within one timestep (see \cref{eq:assumption-standard,eq:assumption} 
in the next section) is different in a standard PSATD scheme with moving
window and in a Galilean PSATD scheme.

\section{\label{sec:cartesian}The Galilean PSATD scheme in Cartesian geometry}

While the previous section gave an intuitive description of the Galilean
scheme, in the present section we introduce the exact numerical
scheme that corresponds to this intuitive description -- in the case of Cartesian geometry.

We start by deriving the update equations for the fields (\cref{sec:deriv-cartesian}). The
resulting PIC loop is then briefly described in
\cref{sec:scheme-cartesian}. Finally, in section
\cref{sec:stability-cartesian}, we show empirically that this PIC scheme
has better stability property than the standard PSATD in the case of a relativistic plasma.

\subsection{\label{sec:deriv-cartesian}Derivation of the discretized Maxwell equations in the Galilean PSATD scheme}

In the PSATD scheme, the Maxwell equations are advanced by tranforming
the fields $\vec{E}$ and $\vec{B}$ into Fourier space, and then by
integrating the Maxwell equations \emph{analytically} over one
timestep. 

In the case of the Galilean PSATD scheme, in order to analytically
integrate the Maxwell equations in \emph{Galilean coordinates}
\cref{eq:maxwell1,eq:maxwell2}, we first decouple the equations for $\vec{E}$ and $\vec{B}$ by combining
\cref{eq:maxwell1,eq:maxwell2} into second-order differential
equations:
\begin{subequations}
\begin{align}
\left( \Dt{\;} -\vgal\cdot\nab\right)^2 \vec{B} - &c^2\nab^2 \vec{B}
  = \frac{1}{\epsilon_0}\nab\times\vec{j} \\
\left( \Dt{\;} -\vgal\cdot\nab\right)^2 \vec{E} - &c^2\nab^2 \vec{E}
  = -\frac{c^2}{\epsilon_0}\nab\rho \nonumber \\
 & - \frac{1}{\epsilon_0}\left(\Dt{\;} -\vgal\cdot\nab\right)\vec{j}
\end{align}
\end{subequations}
Note that we used the equations $\nab\cdot\vec{E} = \rho/\epsilon_0$,
$\vec{\nabla'}\cdot\vec{B}=0$ and $\epsilon_0\mu_0 c^2 = 1$ in order 
to obtain the above equations.

In Fourier space, these equations become:
\begin{subequations}
\begin{align}
\left( \Dt{\;} -i\vec{k}\cdot\vgal\right)^2 \vec{\mc{B}} + &c^2\vec{k}^2 \vec{\mc{B}}
 = \frac{1}{\epsilon_0}i\vec{k}\times\vec{\mc{J}} \label{eq:spectral-2ndorder1}\\
\left( \Dt{\;} - i\vec{k}\cdot\vgal\right)^2 \vec{\mc{E}} + &c^2\vec{k}^2 \vec{\mc{E}}
 = -\frac{c^2}{\epsilon_0}\mc{\rho} \,i\vec{k}\nonumber \\
 & - \frac{1}{\epsilon_0}\left(\Dt{\;}
   -i\vec{k}\cdot\vgal\right)\vec{\mc{J}} \label{eq:spectral-2ndorder2}
\end{align}
\end{subequations}
where the Fourier components are defined by $\mc{F}(\vec{k}, t) = \int
d^3\vec{x}' \,F(\vec{x}', t) \,e^{-i\vec{k}\cdot\vec{x}'} $, and $F$ is
either $\vec{E}$, $\vec{B}$, $\vec{j}$ or $\rho$.

\cref{eq:spectral-2ndorder1,eq:spectral-2ndorder2} are linear ordinary
differential equations in $t$, and they can be integrated
analytically over one timestep (i.e. from $t=n\Delta t$ to
$t=(n+1)\Delta t$), provided that
the time evolutions of the source terms $\vec{\mc{J}}(\vec{k}, t)$ and
$\mc{\rho}(\vec{k}, t)$ are known over this timestep -- or,
equivalently, provided that the time evolutions of the corresponding
real-space function $\vec{j}(\vec{x}', t)$ and $\rho(\vec{x}', t)$ are
known.

However, in a PIC code, $\vec{j}(\vec{x}', t)$ and $\rho(\vec{x}', t)$
are obtained from the deposition of the macroparticles' charge and
current onto the grid. For this reason $\vec{j}(\vec{x}', t)$ and
$\rho(\vec{x}', t)$ are only known at a few discrete times between
$n\Delta t$ and $(n+1)\Delta t$. For instance, in a typical PSATD PIC
cycle, $\vec{j}$ is only computed at time $(n+1/2)\Delta t$ and $\rho$ at
times $n\Delta t$ and $(n+1)\Delta t$.

Thus, in order to analytically integrate equations
\cref{eq:spectral-2ndorder1,eq:spectral-2ndorder2}, one needs
to make explicit assumptions on the time evolution of $\vec{j}$ and $\rho$
between these known times. In the standard PSATD scheme (i.e. when the
Maxwell equations are integrated in the standard coordinates $\vec{x}$), one
typically assumes the current $\vec{j}$ to be constant over one timestep:
\begin{equation}
\label{eq:assumption-standard}
\vec{j}( \vec{x}, t ) = \vec{j}( \, \vec{x}, (n+1/2)\Delta t \,)  \qquad
\forall t \in [ \; n\Delta t, \,(n+1)\Delta t \; ]
\end{equation}
However, when integrating the Maxwell equations in the Galilean
variables $\vec{x}'$, it is more natural to assume
\begin{equation}
\label{eq:assumption}
\vec{j}( \vec{x}', t ) = \vec{j}( \, \vec{x}', (n+1/2)\Delta t \,)  \qquad
\forall t \in [ \; n\Delta t, \,(n+1)\Delta t \; ]
\end{equation}
i.e. that the current is constant over one timestep \emph{in the
  Galilean coordinates}. Because of the definition of $\vec{x}'$ (see \cref{eq:change-var}), the
assumptions \cref{eq:assumption-standard} and \cref{eq:assumption} are
not equivalent. In fact, assuming \cref{eq:assumption} instead of
\cref{eq:assumption-standard} is one of the \emph{key difference} between the 
Galilean PSATD scheme described here, and the standard PSATD scheme.

Once we adopt the assumption \cref{eq:assumption}, our numerical
scheme is fully determined. \cref{eq:assumption} indeed results in:
\begin{equation}
\label{eq:J-evolution}
\vec{\mc{J}}( \vec{k}, t ) = \vec{\mc{J}}(\vec{k}, (n+1/2)\Delta t ) \qquad
\forall t \in [ \; n\Delta t, \,(n+1)\Delta t \; ]
\end{equation}
This equation in turn allows us to infer the time evolution of
$\mc{\rho}(\vec{k}, t)$ between $n\Delta t$ and $(n+1)\Delta t$. Indeed, in
the Galilean coordinates, the equation of continuity reads $\left( \partial_t - \vgal\cdot\nab \right)\rho +
\nab\cdot \vec{j} = 0$, which becomes in Fourier space $\left( \partial_t - i\vec{k}\cdot\vgal \right)\mc{\rho} +
i\vec{k}\cdot \vec{\mc{J}}= 0$. The solution $\mc{\rho}(\vec{k}, t)$ of
this equation for a constant $\vec{\mc{J}}$ is necessarily of the form:
\begin{align}
\label{eq:rho-evolution}
\mc{\rho}(\vec{k}, t) &= \mc{\rho}(\vec{k}, (n+1)\Delta t) \frac{1 -
  e^{i\vec{k}\cdot\vgal(t-n\Delta t)}}{1-e^{i\vec{k}\cdot\vgal\Delta t}} \nonumber \\
& - \mc{\rho}(\vec{k}, n\Delta t) \frac{e^{i\vec{k}\cdot\vgal\Delta t} -
  e^{i\vec{k}\cdot\vgal(t-n\Delta t)}}{1-e^{i\vec{k}\cdot\vgal\Delta t}} 
\end{align}
where we explicitly ensured that this solution satisfies the known initial and final conditions $\mc{\rho}(\vec{k}, n\Delta t)$ and $\mc{\rho}(\vec{k}, (n+1)\Delta t)$, which,
again, are typically obtained from charge deposition during the PIC
cycle. As a side note, notice that a necessary and sufficient
condition for \cref{eq:rho-evolution} to be a solution of the
continuity equation with \cref{eq:J-evolution}
is that the following relation be satisfied:
\begin{equation}
\label{eq:continuity}
-i(\vec{k}\cdot\vgal) \frac{ \mc{\rho}^{n+1}  -
 \mc{\rho}^n e^{i\vec{k}\cdot\vgal\Delta t} }{1-
  e^{i\vec{k}\cdot\vgal\Delta t}} + i\vec{k}\cdot\vec{\mc{J}}^{n+1/2}= 0
\end{equation}
where we introduced the short-hand notations $\mc{\rho}^n \equiv
\mc{\rho}(\vec{k}, n\Delta t)$, $\mc{\rho}^{n+1} \equiv \mc{\rho}(\vec{k},
(n+1)\Delta t)$ and $\vec{\mc{J}}^{n+1/2} \equiv \vec{\mc{J}}(\vec{k}, (n+1/2)\Delta t)$.
Thus \cref{eq:continuity} is the discrete equation that $\mc{\rho}^{n+1}$,
$\mc{\rho}^n$, $\vec{\mc{J}}^{n+1/2}$ should satisfy in order to satisfy 
the continuity equation -- and therefore to ensure charge conservation -- in the
Galilean coordinates. (Notice
that in the limit $\vgal = 0$ this equation reduces to
$(\mc{\rho}^{n+1} - \mc{\rho}^{n})/\Delta t +
i\vec{k}\cdot\vec{\mc{J}}^{n+1/2} = 0$.) As such, it is also the
equation that should be enforced during a PIC cycle, either through an Esirkepov-type
deposition scheme or through a current correction scheme.

Finally, the time evolution of $\mc{\rho}$ and $\vec{\mc{J}}$
(\cref{eq:J-evolution,eq:rho-evolution}) is inserted into the
right-hand side of the Maxwell equations
\cref{eq:spectral-2ndorder1,eq:spectral-2ndorder2}. Again, these
equations are linear ordinary differential equations, now with explicit
expressions in their right-hand side, and they can be integrated analytically. Integrating these equations from $t=n\Delta
t$ to $t=(n+1)\Delta t$ results in the following update equations (see
appendix \ref{app:Maxwell-integration} for the details of the derivation):
\begin{subequations}
\begin{align}
\vec{\mc{B}}^{n+1} &= \theta^2 C \vec{\mc{B}}^n
 -\frac{\theta^2 S}{ck}i\vec{k}\times \vec{\mc{E}}^n
+ \;\frac{\theta \chi_1}{\epsilon_0c^2k^2}\;i\vec{k} \times
                     \vec{\mc{J}}^{n+1/2} \label{eq:disc-maxwell1}\\
\vec{\mc{E}}^{n+1} &=  \theta^2 C  \vec{\mc{E}}^n
 +\frac{\theta^2 S}{k} \,c i\vec{k}\times \vec{\mc{B}}^n 
+\frac{i\nu \theta \chi_1 - \theta^2S}{\epsilon_0 ck} \; \vec{\mc{J}}^{n+1/2}\nonumber \\
& - \frac{1}{\epsilon_0k^2}\left(\; \chi_2\;\mc{\rho}^{n+1} -
  \theta^2\chi_3\;\mc{\rho}^{n} \;\right) i\vec{k} \label{eq:disc-maxwell2}
\end{align}
\end{subequations}
where we used the short-hand notations $\vec{\mc{E}}^n \equiv
\vec{\mc{E}}(\vec{k}, n\Delta t)$, $\vec{\mc{B}}^n \equiv
\vec{\mc{B}}(\vec{k}, n\Delta t)$ as well as:
\begin{subequations}
\begin{align}
&C = \cos(ck\Delta t) \quad S = \sin(ck\Delta t) \quad k
= |\vec{k}| \label{eq:def-C-S}\\&
\nu = \frac{\vec{k}\cdot\vgal}{ck} \quad \theta =
  e^{i\vec{k}\cdot\vgal\Delta t/2} \quad \theta^* =
  e^{-i\vec{k}\cdot\vgal\Delta t/2} \label{eq:def-nu-theta}\\&
\chi_1 =  \frac{1}{1 -\nu^2} \left( \theta^* -  C \theta + i
  \nu \theta S \right) \label{eq:def-chi1}\\&
\chi_2 = \frac{\chi_1 - \theta(1-C)}{\theta^*-\theta} \quad
\chi_3 = \frac{\chi_1-\theta^*(1-C)}{\theta^*-\theta} \label{eq:def-chi23}
\end{align}
\end{subequations}
Note that, in the limit $\vgal=\vec{0}$,
\cref{eq:disc-maxwell1,eq:disc-maxwell2} reduce to the standard PSATD
equations \cite{Haber}, as expected.

\subsection{\label{sec:scheme-cartesian}Overview of the PIC cycle for the Galilean PSATD scheme}

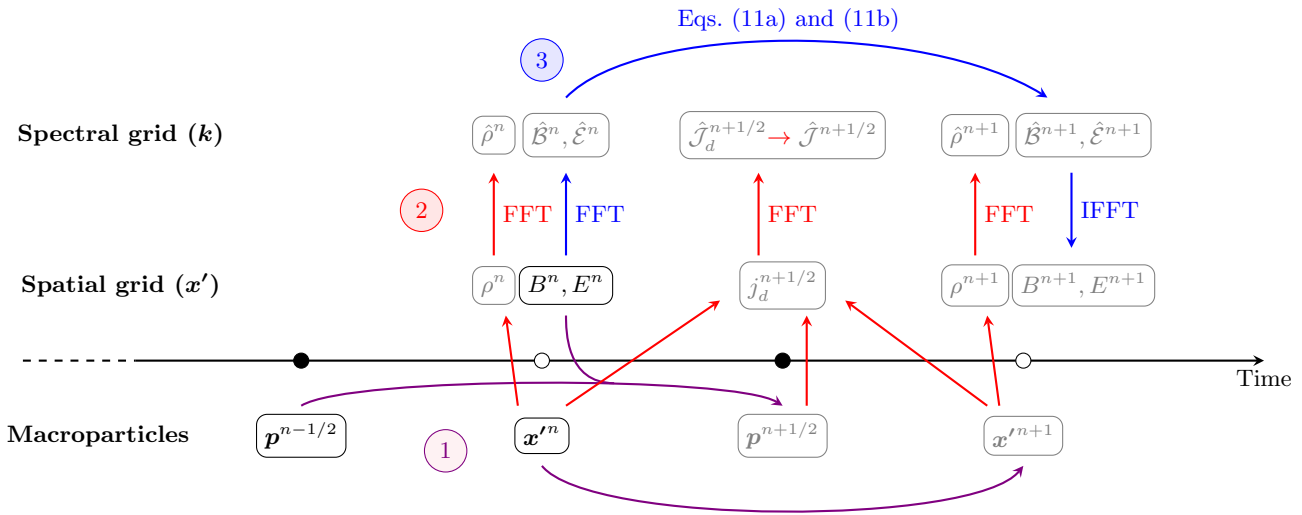
\begin{figure*}
\begin{tikzpicture}

\def \Dt{3.2}
\def \yspac{1.}

\draw[thick,->,>=stealth] (1,0) -- (5*\Dt,0) node[below]{Time};
\draw[thick,dashed] (-0.5,0) -- (1,0);

\draw (0.8,3*\yspac) node{\textbf{Spectral grid ($\vec{k}$)}};
\draw (0.8,\yspac) node{\textbf{Spatial grid ($\vec{x'}$)}};
\draw (0.5,-\yspac) node{\textbf{Macroparticles}};
\foreach \n in {1,3} 
\draw[fill=black] (\n*\Dt,0) circle(0.1);
\foreach \n in {2,4} 
\draw[fill=white] (\n*\Dt,0) circle(0.1);

\draw (\Dt,-\yspac) node[black,fill=white,draw=black,rounded corners]{$\vec{p}^{n-1/2}$};
\draw (1.8*\Dt,\yspac) node[gray,draw=gray,rounded corners]{$\rho^n$};
\draw (2.1*\Dt,\yspac) node[black,draw=black,rounded corners]{$B^n, E^n$};
\draw (1.8*\Dt,3*\yspac) node[gray,fill=white,draw=gray,rounded corners]{$\mc{\rho}^n$};
\draw (2.1*\Dt,3*\yspac) node[gray,fill=white,draw=gray,rounded corners]{$\mc{B}^n, \mc{E}^{n}$};
\draw (2*\Dt,-\yspac) node[black,fill=white,draw=black,rounded corners]{$\vec{x'}^{n}$};
\draw (3*\Dt,3*\yspac) node[gray,fill=white,draw=gray,rounded corners]{$ \mc{J}_d^{n+1/2} \textcolor{red}{\rightarrow} \;\mc{J}^{n+1/2}$};
\draw (3*\Dt,1*\yspac) node[gray,fill=white,draw=gray,rounded corners]{$ j_d^{n+1/2}$};
\draw (3*\Dt,-\yspac) node[gray,fill=white,draw=gray,rounded corners]{$\vec{p}^{n+1/2}$};
\draw (3.8*\Dt,3*\yspac) node[gray,fill=white,draw=gray,rounded corners]{$\mc{\rho}^{n+1}$};
\draw (4.25*\Dt,3*\yspac) node[gray,fill=white,draw=gray,rounded corners]{$  \mc{B}^{n+1}, \mc{E}^{n+1}$};
\draw (3.8*\Dt,\yspac) node[gray,fill=white,draw=gray,rounded corners]{$\rho^{n+1}$};
\draw (4.25*\Dt,\yspac) node[gray,fill=white,draw=gray,rounded corners]{$B^{n+1}, E^{n+1}$};
\draw (4*\Dt,-\yspac) node[gray,fill=white,draw=gray,rounded corners]{$\vec{x'}^{n+1}$};

\draw[->,blue!50!red,>=stealth,thick] (1*\Dt,-0.6*\yspac) .. controls (1.2*\Dt,-0.2*\yspac) and (2.8*\Dt,-0.2*\yspac) .. (3*\Dt,-0.6*\yspac);
\draw[->,blue!50!red,>=stealth,thick] (2*\Dt,-1.4*\yspac) .. controls (2.2*\Dt,-2.2*\yspac) and (3.8*\Dt,-2.2*\yspac) .. (4*\Dt,-1.4*\yspac);
\draw (1.6*\Dt,-1.2*\yspac) node[blue!50!red,circle,draw=blue!50!red,fill=blue!5!red!5!white]{1};
\draw[blue!50!red,thick] (2.1*\Dt,0.6*\yspac) .. controls
(2.1*\Dt,0.4*\yspac) and (2.1*\Dt,-0.3*\yspac) .. (2.3*\Dt,-0.3*\yspac);

\draw[->,red,>=stealth,thick] (2.1*\Dt,-0.6*\yspac) -- (2.74*\Dt,0.8*\yspac);
\draw[->,red,>=stealth,thick] (3.1*\Dt,-0.6*\yspac) -- (3.1*\Dt,0.6*\yspac);
\draw[->,red,>=stealth,thick] (2.9*\Dt,1.4*\yspac) -- node[red, anchor=west]{FFT}(2.9*\Dt,2.5*\yspac);
\draw[->,red,>=stealth,thick] (3.85*\Dt,-0.6*\yspac) -- (3.26*\Dt,0.8*\yspac);
\draw[->,red,>=stealth,thick] (3.9*\Dt,-0.6*\yspac) -- (3.85*\Dt,0.6*\yspac);
\draw[->,red,>=stealth,thick] (1.9*\Dt,-0.6*\yspac) -- (1.85*\Dt,0.6*\yspac);
\draw[->,red,>=stealth,thick] (3.8*\Dt,1.4*\yspac) -- node[red, anchor=west]{FFT}(3.8*\Dt,2.5*\yspac);
\draw[->,red,>=stealth,thick] (1.8*\Dt,1.4*\yspac) -- node[red, anchor=west]{FFT}(1.8*\Dt,2.5*\yspac);
\draw (1.5*\Dt,2*\yspac) node[red,circle,draw=red,fill=red!10!white]{2};

\draw[->,blue,>=stealth,thick] (2.1*\Dt,1.4*\yspac) -- node[anchor=west,blue]{FFT}(2.1*\Dt,2.5*\yspac);
\draw[<-,blue,>=stealth,thick] (4.2*\Dt,1.5*\yspac) -- node[anchor=west,blue]{IFFT}(4.2*\Dt,2.5*\yspac);
\draw[->,blue,>=stealth,thick] (2.1*\Dt,3.5*\yspac) .. controls (2.4*\Dt,4.5*\yspac) and  (3.6*\Dt,4.5*\yspac) .. node[above,blue]{\cref{eq:disc-maxwell1,eq:disc-maxwell2}} (4.1*\Dt,3.5*\yspac);
\draw (2*\Dt,4*\yspac) node[blue,circle,draw=blue,fill=blue!10!white]{3};

\end{tikzpicture}
\caption{\label{fig:cycle}Schematic representation of the PIC
  cycle. The quantities that are known at the beginning of the PIC
  cycle are displayed in black, while the quantities that are
  computed during the PIC cycle are displayed in gray. The three
  successive steps of the PIC cycle -- particle push (1), current
  deposition (2) and Maxwell solver (3) -- are represented in purple,
  red and blue respectively.}
\end{figure*}

\cref{eq:continuity,eq:disc-maxwell1,eq:disc-maxwell2} are the
fundamental field equations of our PIC cycle. While \cref{sec:deriv-cartesian} emphasized the logical reasoning that leads to
these equations, it did not give a precise description of their role
within the PIC cycle. Therefore, the present section gives a concise overview of the different steps of the PIC cycle, for the Galilean PSATD scheme.

Apart from the fact that the simulation is performed in Galilean
coordinates, our PIC cycle is very close to the standard PSATD scheme
\cite{Haber}. In particular, the fields $\vec{E}$, $\vec{B}$ and
$\rho$ and the macroparticles' positions $\vec{x}'$ are defined at integer
times, whereas the field $\vec{J}$ and the macroparticles' momenta
$\vec{p}$ are defined at half-integer times. All the fields are
defined at the same points
in space (i.e. the spatial grid is not staggered). The successive steps
of the PIC cycle are represented in \cref{fig:cycle} and described below.

\subsubsection{Particle push} 
The fields $\vec{E}$ and $\vec{B}$ are interpolated at time $t=n\Delta
t$ from the spatial grid to the
macroparticles' positions (naturally using the Galilean coordinates $\vec{x}'$ for
the interpolation). The interpolated fields are then used to push the
macroparticles' momenta from $t = (n-1/2)\Delta t$ to $t=(n+1/2)\Delta
t$, using a discretized version of the equation of motion
\cref{eq:motion2}. Note that \cref{eq:motion2} is familiar, and can be discretized by
using e.g. the Boris pusher \cite{Boris1970} or, as we chose in this paper, the
Vay pusher \cite{VayPoP2008}. Then the macroparticles' positions are pushed from $t = n\Delta
t$ to $t=(n+1)\Delta t$ by using a trivial leap-frog discretization of
\cref{eq:motion1}:
\begin{equation}
\label{eq:disc-motion}
\vec{x}'^{n+1} = \vec{x}'^n + \Delta t\left(
  \frac{\vec{p}^{n+1/2}}{\gamma^{n+1/2} m} - \vgal \right)
\end{equation}
where $\gamma^{n+1/2} = \sqrt{1 + ( \vec{p}^{n+1/2} / mc )^2 }$.

\subsubsection{Current and charge deposition} 
The charge density $\rho$
is then computed on the spatial grid at $t=n\Delta t$ and
$t=(n+1)\Delta t$ from the macroparticles' positions $\vec{x}'^n$ and
$\vec{x}'^{n+1}$ respectively. In addition, by using the intermediate
positions $\vec{x}'^{n+1/2}\equiv
(\vec{x}'^n+\vec{x}'^{n+1})/2$, the current $\vec{j}_d$ is calculated on the
spatial grid at $t=(n+1/2)\Delta t$. Here, the subscript $d$ emphasizes
the fact that we use a \emph{direct} deposition scheme, rather than a 
charge-conserving deposition scheme. As a consequence, the
Fourier transform $\vec{\mc{J}}^{n+1/2}_d$ of the current
$\vec{j}_d^{n+1/2}$ does not satisfy the discretized continuity
equation \cref{eq:continuity} by default. For this reason, we use
the Fourier transform of the charge density at $t=n\Delta t$ and
$t=(n+1)\Delta t$ to compute a corrected current
$\vec{\mc{J}}^{n+1/2}$ which does satisfy \cref{eq:continuity}:
\begin{subequations}
\begin{align}
&\vec{\mc{J}}^{n+1/2} = \vec{\mc{J}}_d^{n+1/2} +
\frac{i\vec{k}}{k^2}\vec{\mc{G}} \label{eq:current-correction}\\ 
&\vec{\mc{G}} = -i(\vec{k}\cdot\vgal) \frac{ \mc{\rho}^{n+1}  -
 \mc{\rho}^n e^{i\vec{k}\cdot\vgal\Delta t} }{1-
  e^{i\vec{k}\cdot\vgal\Delta t}} + i\vec{k}\cdot\vec{\mc{J}}_d^{n+1/2}
\end{align}
\end{subequations}

Finally, a light amount of spatial smoothing is applied to
$\mc{\rho}^n$, $\mc{\rho}^{n+1}$ and $\vec{\mc{J}}^{n+1/2}$. More
precisely, each of these fields is multiplied by a low-pass filter
which is equivalent, in real-space, to a one-pass binomial smoother followed by a compensator \cite{Birdsall2004}:
\begin{align}
\mc{T}(\vec{k}) =& \left(1-\sin^2(k_x \Delta x/2)\right)\left( 1 + \sin^2(k_x \Delta
                   x/2)\right) \nonumber \\
  & \times \left(1-\sin^2(k_y \Delta y/2)\right)\left( 1 + \sin^2(k_y \Delta
    y/2)\right) \nonumber \\
& \times \left(1-\sin^2(k_z \Delta z/2)\right)\left( 1 + \sin^2(k_z \Delta
    z/2)\right) \label{eq:smoothing}
\end{align}
where $\Delta x$, $\Delta y$, $\Delta z$ are the cell size of the spatial
grid in each direction. (For 2D simulations in the $x$-$z$ plane, this
expression is applied with $k_y=0$.)

\subsubsection{Maxwell solver} 

In order to update the values of the fields $\vec{E}$ and $\vec{B}$
from $t=n\Delta t$ to $t=(n+1)\Delta t$, we first transform them to
Fourier space at $t=n\Delta t$. We then use the deposited fields
$\mc{\rho}^n$, $\mc{\rho}^{n+1}$ and $\vec{J}^{n+1/2}$ as well as
\cref{eq:disc-maxwell1,eq:disc-maxwell2} to obtain the updated values
$\vec{\mc{E}}^{n+1}$ and $\vec{\mc{B}}^{n+1}$ in spectral
space. Finally, these fields are converted back to real space by using
an inverse Fourier transform.

\subsection{\label{sec:stability-cartesian}Stability of a uniform, relativistic plasma}

We implemented the Galilean PSATD scheme described in
\cref{sec:scheme-cartesian} in the code Warp \cite{Warpref}. We then tested its
stability for simulations of relativistic flowing plasmas, in a 2D
Cartesian geometry.

\begin{figure}
\includegraphics[width=\columnwidth]{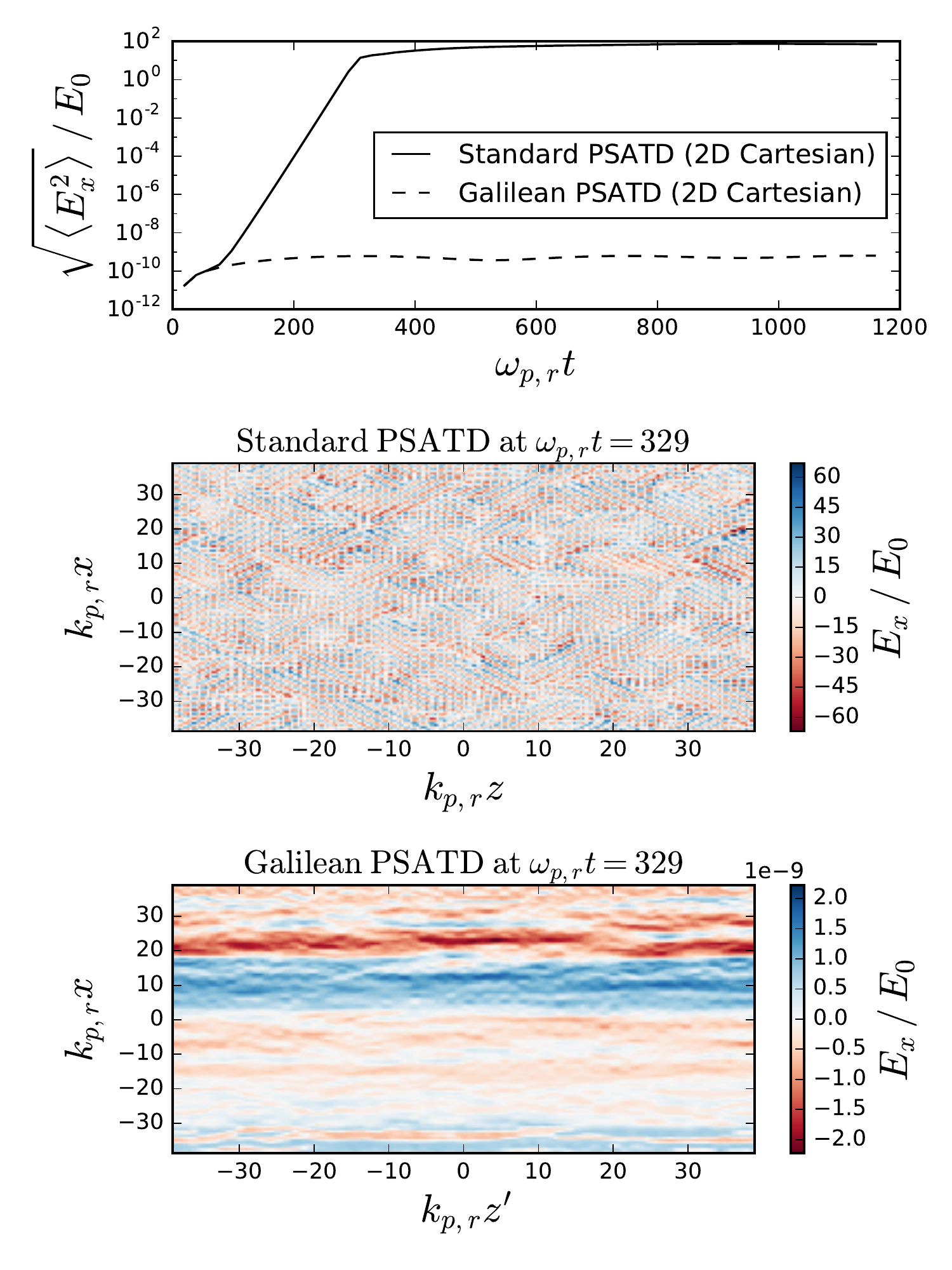}
\caption{\label{fig:empirical}Results of a relativistic flowing plasma
simulation in Warp, with the standard and Galilean PSATD scheme. 
Top panel: RMS amplitude of the
electric field in the simulation box versus time. (By
definition, $\omega_{p,r} \equiv \omega_p/\gamma_0^{1/2}$ and $E_0
\equiv m_e c \omega_{p,r}/e$.) Middle and bottom panels: Maps
of the electric field in the simulation box, at a given time, for
both the standard and Galilean PSATD scheme. ($k_{p,r} \equiv k_p/\gamma_0^{1/2}$)}
\end{figure} 

\begin{table}
\caption{\label{tab:parameters}Parameters of the test simulations. The simulations are scaled by $k_{p,r} \equiv
  k_p/\gamma_0^{1/2}$, with $k_p^2 = n_0 e^2/m_e \epsilon_0 c^2$.}
\begin{ruledtabular}
\begin{tabular}{r l}
Plasma density & $n_0$ (scales the simulation) \\
Lorentz factor & $\gamma_0 = 130$ \\
Cell size along $z$ & $\Delta z = 0.3868 \; k_{p,r}^{-1}$ \\
Cell size along $x$ & $\Delta x = 0.3868 \; k_{p,r}^{-1}$ \\
Timestep & $\Delta t = \Delta z/c$ \\
Number of gridpoints & $N_x = N_z = 200$ \\
Order of the shape factor & 3 (both in $x$ and $z$)
\end{tabular}
\end{ruledtabular}
\end{table}
In the test simulations, a uniform plasma of density $n_0$ 
fills a periodic simulation box, and flows towards the positive $z$ with a
relativistic speed. The physical and numerical parameters of the
simulation are summarized in \cref{tab:parameters}. We ran this
simulation both with the standard PSATD scheme and with the Galilean
PSATD scheme, using $\vgal = \vec{v}_0$ in the latter
case, with $\vec{v}_0$ the velocity of the plasma. 
(Thus in the standard PSATD simulation, the relativistic
plasma cycles through the fixed periodic boundaries of the box, while
in the Galilean PSATD simulation, the box moves along with the plasma.) 

The results of this test are shown in \cref{fig:empirical}. As shown
in the top panel, in the case of the standard PSATD the fluctuations
of the electric fields grow exponentially and saturate shortly after the
beginning of the simulation. This a result of the well-known NCI. 
Conversely, with the Galilean PSATD the
fluctuations of $E_x$ remain at a low level, and can be explained by
a simple accumulation of numerical noise. This interpretation is
confirmed by the maps of the electric field in the middle and bottom
panels. While the standard PSATD simulation exhibits a high-wavenumber
pattern that is characteristic of the NCI, the
Galilean PSATD simulation exhibits a random-looking pattern (with an
amplitude that is lower by almost 10 orders of magnitude) that is
consistent with numerical or thermal noise.

The Galilean PSATD scheme is thus empirically much more stable than
the standard PSATD scheme. Again, a remarkable point is that we did
not introduce any NCI-specific correction here. 
Instead, the Galilean scheme simply results from the natural
analytical integration of the Maxwell equations in the Galilean
coordinates, with no additional corrections.

\section{\label{sec:stability-analysis}Stability analysis in 2D 
Cartesian geometry}

While the previous section showed \emph{empirically} that the Galilean
PSATD scheme is more stable, in the present section we confirm and
explain these results by using the theoretical dispersion relation that
corresponds to this numerical scheme.

\subsection{Dispersion equation}

More precisely, we start with a neutral, uniform plasma, flowing with a
velocity $\vec{v}_0 = v_0\vec{u}_z$ (and Lorentz factor $\gamma_0$)
through a 2D periodic grid, and we consider the evolution of a small
perturbation to its fields, of the form: 
\begin{equation}
\vec{E}, \vec{B} \propto e^{i\vec{k}\cdot\vec{x} - i\omega t} =
e^{i\vec{k}\cdot\vec{x}' - (\omega - \vec{k}\cdot\vgal)t}
\end{equation}
Notice that, with the above definition, the physical interpretation of
$\omega$ and $\vec{k}$ is the natural one, and in particular this interpretation does
not depend on the choice of $\vgal$.

By combining the perturbed Vlasov equation and the Maxwell equations, we
obtain a dispersion equation that relates $\omega$ and
$\vec{k}$. Importantly, the analysis -- and the resulting dispersion
equation -- incorporate all the numerical effects that are introduced
by the PIC cycle from \cref{sec:scheme-cartesian} (including finite
timestep, finite spatial resolution, shape factors, current
correction, etc.). Note however that the analysis has been restricted
to the case where $\vgal$ is along $z$ (i.e. $\vgal = v_{gal}\vec{u}_z$). The full derivation of the dispersion
equation is given in appendix \ref{sec:derivation-dispersion}. Although
this derivation builds upon previous work
\cite{GodfreyJCP1975,GodfreyJCP2013,XuCPC2013,GodfreyJCP2014}, a number of important changes have been introduced in order to accomodate the
specifics of the Galilean PSATD scheme.

The resulting dispersion relation is given in \cref{eq:dispersion}, along
with the expression of the dimensionless coefficients $\xi$
(\cref{eq:xi1,eq:xi2,eq:xi3}), which represent the response of the
plasma, and include the effects of spatial field smoothing
($\mc{T}(\vec{k})$), spatial aliases ($\Km = m_x \frac{2\pi}{\Delta x}\vec{u}_x + m_z \frac{2\pi}{\Delta z}\vec{u}_z$) and finite shape factor ($\mc{S}(\vec{k})$). The factor
$\mc{S}(\vec{k})$ indeed represents the Fourier transform of
the macroparticle shape factor, so that e.g. for a shape factor of order
$\ell_x $ and $\ell_z $ along $x$ and $z$ respectively, one has:
\begin{equation}
\mc{S}(\vec{k}) = \mathrm{sinc}^{\ell_x + 1}\left( \frac{k_x\Delta
    x}{2} \right) \mathrm{sinc}^{\ell_z + 1}\left( \frac{k_z\Delta z}{2} \right)
\end{equation}
with $\mathrm{sinc}(x) = \sin(x)/x$. In addition, in
\cref{eq:dispersion,eq:chi5} we also used the short-hand notations
\begin{equation}
s_x = \sin\left(\frac{x\Delta t}{2}\right) \quad c_x =
\cos\left(\frac{x\Delta t}{2}\right) \quad t_x = \tan\left(\frac{x\Delta t}{2}\right)
\end{equation}
and we introduced $\chi_5$ and $\chi_5'$, which are 
coefficients that depend only on $\omega$, $k$ and $\nu = \vec{k}\cdot
\vgal/(ck)$, and whose mathematical expression results from the key
hypothesis \cref{eq:assumption}.
\begin{widetext}
\begin{align}
\label{eq:dispersion}
&(s_\omega^2 - t_{ck}^2 c_\omega^2  )\left[ 1 - \frac{1}{k}\left( k_x\xi_{3x} +
    \frac{k_z\xi_{3z}}{\gamma_0^2} \right) \right]  - \xi_1 \left[
  \frac{\chi_5}{k^2} \left( k_z^2 + \frac{k_x^2}{\gamma_0^2} \right) +
  \chi_5' \frac{k_z v_0}{ck} \right] 
 \nonumber \\
&+ \frac{k_xv_0}{ck}\left[
  \frac{\chi_5}{k}\left( k_z\xi_{2x} - \frac{k_x\xi_{2z}}{\gamma_0^2}
  \right) + \chi_5'\frac{\xi_{2x}v_0}{c} \right] 
+ \frac{1}{\gamma_0^2}\left(\chi_5 + \frac{k_z v_0}{ck}
  \chi_5'\right)\left[ \xi_1\frac{k_x\xi_{3x} + k_z\xi_{3z}}{k} 
+ \frac{k_x v_0}{ck}(\xi_{3x}\xi_{2z} - \xi_{3z}\xi_{2x})\right] = 0
\end{align}
\begin{subequations} 
with
\begin{equation}
 \xi_1 =\mc{T}(\vec{k})  \frac{\omega_p^2}{\gamma_0 c k} \sum_{m_x,
         m_z=-\infty}^{\infty}\frac{1}{\frac{2}{\Delta t}\sin\left[ (\omega-k_z
         v_0)\frac{\Delta t}{2} + m_z\frac{\pi \Delta t}{\Delta
         z}(v_{gal}- v_{0})\right]}  \mc{S}^2(\vec{k} +
         \Km) \label{eq:xi1}
\end{equation}
\begin{equation}
 \vec{\xi}_2 =\mc{T}(\vec{k})  \frac{\omega_p^2}{\gamma_0 k} \sum_{m_x,
         m_z=-\infty}^{\infty}\frac{\cos\left[ (\omega-k_z
         v_0)\frac{\Delta t}{2} + m_z\frac{\pi \Delta t}{\Delta
         z}(v_{gal}- v_{0})\right]}{\left(\frac{2}{\Delta
               t}\right)^2\sin^2\left[ (\omega-k_z
         v_0)\frac{\Delta t}{2} + m_z\frac{\pi \Delta t}{\Delta
         z}(v_{gal}- v_{0})\right]}  \mc{S}^2(\vec{k} +
         \Km)(\vec{k}+\Km) \label{eq:xi2}
\end{equation}
\begin{equation}
 \vec{\xi}_3 =\mc{T}(\vec{k})  \frac{\omega_p^2}{\gamma_0 k} \sum_{m_x,
         m_z=-\infty}^{\infty} \frac{1}{\left(\frac{2}{\Delta
               t}\right)^2\sin^2\left[ (\omega-k_z
         v_0)\frac{\Delta t}{2} + m_z\frac{\pi \Delta t}{\Delta
         z}(v_{gal}- v_{0})\right]}    \mc{S}^2(\vec{k} +
         \Km)(\vec{k}+\Km) \label{eq:xi3}
\end{equation}
\end{subequations}
\begin{equation}
\label{eq:chi5}
\chi_5 = \frac{c_\omega c_{\nu ck}}{1-\nu^2}\left( t_\omega(t_{ck} -
         \nu t_{\nu ck}) - t_{ck}(t_{\nu ck} - \nu t_{ck})\right)
       \qquad \chi_5' = \frac{c_\omega c_{\nu ck}}{1-\nu^2}\left( t_\omega(t_{\nu
         ck} - \nu t_{ck}) - t_{ck}(t_{ck} - \nu t_{\nu ck})\right)
\end{equation}
\end{widetext}

Several remarks can be made on the dispersion equation
\cref{eq:dispersion}. First of all, note that the set of equations
\cref{eq:dispersion,eq:xi1,eq:xi2,eq:xi3,eq:chi5} is valid for any value of $v_{gal}$, including $v_{gal}= 0$ (standard
PSATD) and $v_{gal}=v_0$ (optimal Galilean PSATD).

Another important point is that it
can be verified (although only after some algebra) that
\cref{eq:dispersion} reduces, for any value of $v_{gal}$, to 
\begin{equation}
\frac{\Delta t^2}{4} \times \left( \omega^2 - c^2k^2 -
  \frac{\omega_p^2}{\gamma_0}\right) \times \left( 1 - \frac{\omega_p^2}{\gamma_0^3
    (\omega - k_z v_0)^2} \right) =0
\end{equation}
in the limit of infinitely small timestep and cell size ($\omega
\Delta t\ll 1$, $k\Delta x \ll 1$,  $k\Delta z \ll 1$). Thus, as
expected, in the limit of infinitely high resolution, the dispersion equation
recovers the two independent physical modes of a relativistic plasma -- the
relativistic plasma mode $\omega = k_z v_0 \pm \omega_p \gamma_0^{-3/2}$
and the relativistic electromagnetic mode $\omega^2 = c^2k^2 + \omega_p^2
/\gamma_0$.

Conversely, at finite resolution, \cref{eq:dispersion} gives rise
to distorted modes, which can potentially become unstable. This is
particularly true near the numerical resonances of the plasma coefficients
$\xi_1$, $\vec{\xi}_2$, $\vec{\xi}_3$, i.e. whenever 
\begin{equation}
\label{eq:resonant-condition}
\omega - k_z v_0 + m_z\frac{2\pi}{\Delta z}(v_{gal}- v_0) = 0 \quad
\left(\mathrm{modulo}\; \frac{2\pi}{\Delta t}\right) 
\end{equation}
so that the sine term in the denominators of \cref{eq:xi1,eq:xi2,eq:xi3} goes to 0. Since the resonance condition \cref{eq:resonant-condition} depends on
the alias number $m_z \in \mathbb{Z}$, this equation expresses the well-known
fact that resonances occur at a set of different frequencies (aliased resonances) \cite{GodfreyJCP2013,XuCPC2013,GodfreyJCP2014}.

In this regard, one consequence of the Galilean coordinates is clear: when
choosing $v_{gal} =v_0$, the term proportional to $m_z$ in the
resonance condition \cref{eq:resonant-condition} vanishes, and thus
all the aliased resonances are \emph{relocated} to the same frequency:
$\omega -k_z v_0 = 0$ (modulo $2\pi/\Delta t$). Interestingly, when
tracking the corresponding terms throughout appendix
\ref{sec:derivation-dispersion}, one realizes that this 
\emph{relocation} of resonances is a direct consequence of the fact that the grid 
follows the plasma (as shown in \cref{fig:schematic}), and that it does not
depend on making the assumption \cref{eq:assumption} as opposed to \cref{eq:assumption-standard}.

Aside from this effect, the only other impact of $v_{gal}$ on
the dispersion equation \cref{eq:dispersion} is in the expression of 
the coefficients $\chi_5$ and $\chi_5'$ (through $\nu =
\vec{k}\cdot\vgal/(ck)$), which on the other hand does result from the assumption
\cref{eq:assumption}. This leads us to think that, in the case
$v_{gal} = v_0$, there are special relationships between
$\chi_5$ and $\chi_5'$, which effectively cancel the relocated
resonance. For instance, it can be shown that, in the case
$v_{gal}=v_0$, the factor $(\chi_5+(k_z v_0/ck) \chi_5')$ in the last
term of \cref{eq:dispersion} cancels at the resonance, whereas this is
not true for $v_{gal}=0$. 

Beyond these first remarks, it is difficult to analytically extract more
insights from \cref{eq:dispersion}, and thus this
equation needs to be solved \emph{numerically} in order to actually predict the
stability of a given situation.

\subsection{Numerical solution and comparison with simulations}

\begin{figure*}
\includegraphics[width=0.8\textwidth]{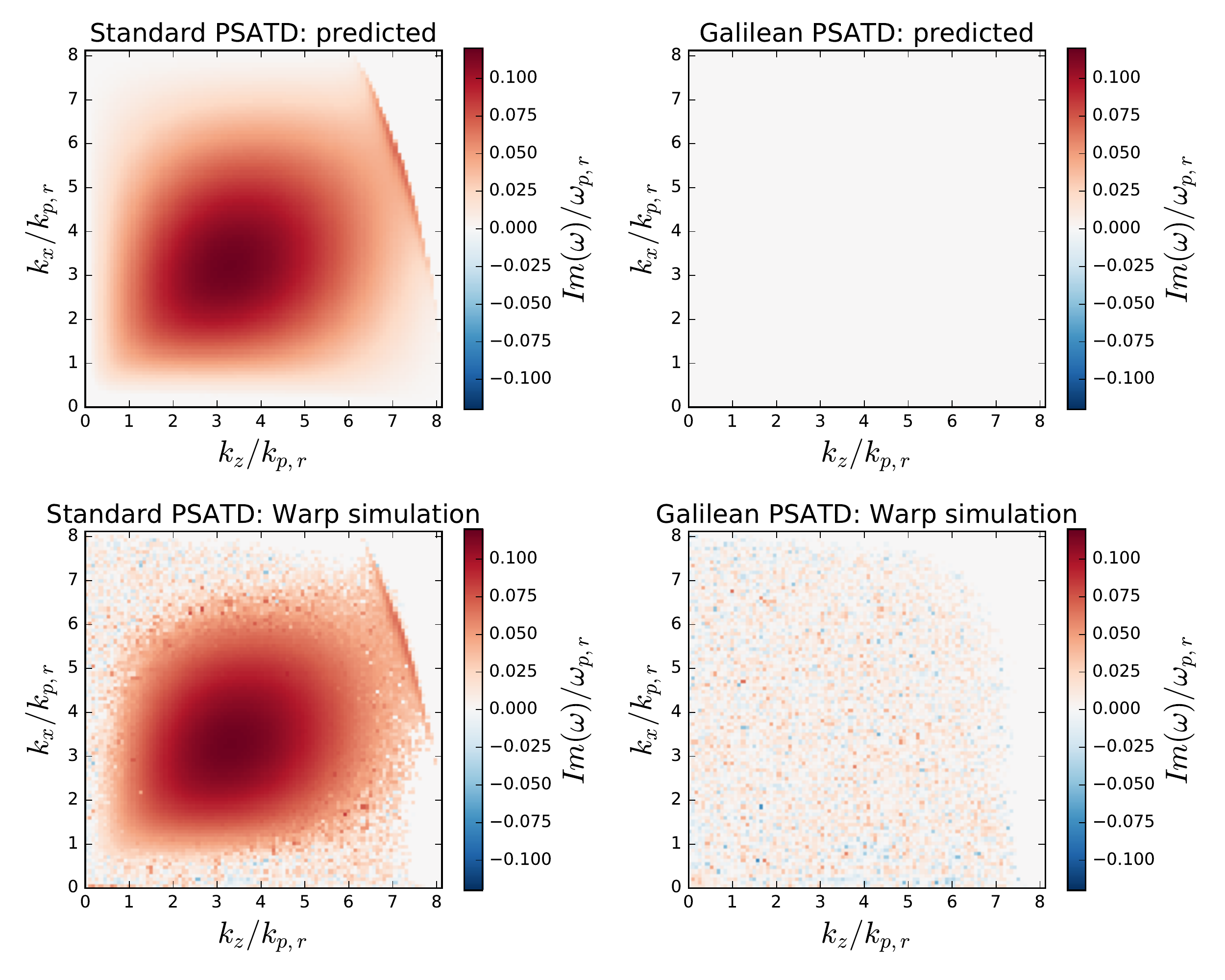}
\caption{\label{fig:predictions}Predicted and observed growth rates
  $\mathrm{Im}(\omega)$ of the NCI, in
  $k$ space (with $k_{p,r} = k_p/\gamma_0^{1/2}$). The predicted
  growth rates (upper panels) are obtained from
  \cref{eq:dispersion}. All four panels use the parameters
  from \cref{tab:parameters}, and in addition the Galilean PSATD (right
  panels) uses $\vgal=v_0\vec{u}_z$.}
\end{figure*}

We solved the dispersion equation \cref{eq:dispersion} numerically, with the
physical and numerical parameters from
\cref{tab:parameters}. In particular, when solving \cref{eq:dispersion} for
$\omega$, we allowed of course for a non-zero imaginary part -- since
$\mathrm{Im}(\omega)$ corresponds the growth rate of the instability.

These predicted growth rates were calculated for $v_{gal} = 0$
(standard PSATD) and $v_{gal} = v_0$ (optimal Galilean PSATD), and
they were compared with the corresponding Warp simulations from
\cref{sec:stability-cartesian}. The results of these comparisons are
shown in \cref{fig:predictions}. Note that in the Warp simulations, the growth
rate was estimated by taking the Fourier transform of the fields at
$\omega_{p,r} t \simeq 120$ and $\omega_{p,r} t\simeq 170$ (i.e. within the
linear growth phase; see \cref{fig:empirical}) and, for each Fourier
mode, by calculating the difference in amplitude between those two
times.

In the case of the standard PSATD (left panels in \cref{fig:predictions}),
one can see that the dispersion equation \cref{eq:dispersion}
correctly predicts that the simulation is unstable (existence of positive,
non-zero $\mathrm{Im}(\omega)$). Moreover, the predicted growth rates
from \cref{eq:dispersion} are in excellent quantitative agreement
with the growth rates observed in the Warp simulation. Notice also
that, in the left panels of \cref{fig:predictions}, the unstable modes 
cluster in two areas of $k$ space: on a fine line
at high $k$, which corresponds to the resonance $m_z=0$ from
\cref{eq:resonant-condition}, and on a broader area at lower $k$. This
second, broader area corresponds to a non-resonant instability, which
has also been predicted and observed in previous work 
\cite{GodfreyJCP2014,GodfreyIEEE2014,GodfreyCPC2015}.

In the case of the Galilean PSATD ($v_{gal} = v_0$; right
panels in \cref{fig:predictions}), the dispersion equation
\cref{eq:dispersion} predicts that all modes are stable
($\mathrm{Im}(\omega)=0$ across all $k$ space). This is again
consistent with the observations from the Warp simulation, since the lower
right panel in \cref{fig:predictions} displays only noise, with both
positive and negative values of $\mathrm{Im}(\omega)$. Again, in our
understanding from the dispersion relation \cref{eq:dispersion}, 
this elimination of both the resonant and non-resonant
NCI is due to mathematical expression of $\chi_5$ and $\chi_5'$, which
result from the assumption on the time evolution of $\vec{j}$ \cref{eq:assumption}.

On the whole, this section confirms that the Galilean PSATD scheme 
eliminates the NCI, since the absence of NCI was both predicted
theoretically and observed in simulations. Remarkably, the Galilean
scheme simultaneously supresses both the high-$k$ resonant instability 
and low-$k$ non-resonant instability. This contrasts with some of the previous
mitigation techniques, which typically introduced two separate numerical corrections
in order to handle the resonant and non-resonant instabilities respectively.

\subsection{\label{sec:influence-vgal}Influence of $v_{gal}$ on the growth rate}

\begin{figure}
\includegraphics[width=\columnwidth]{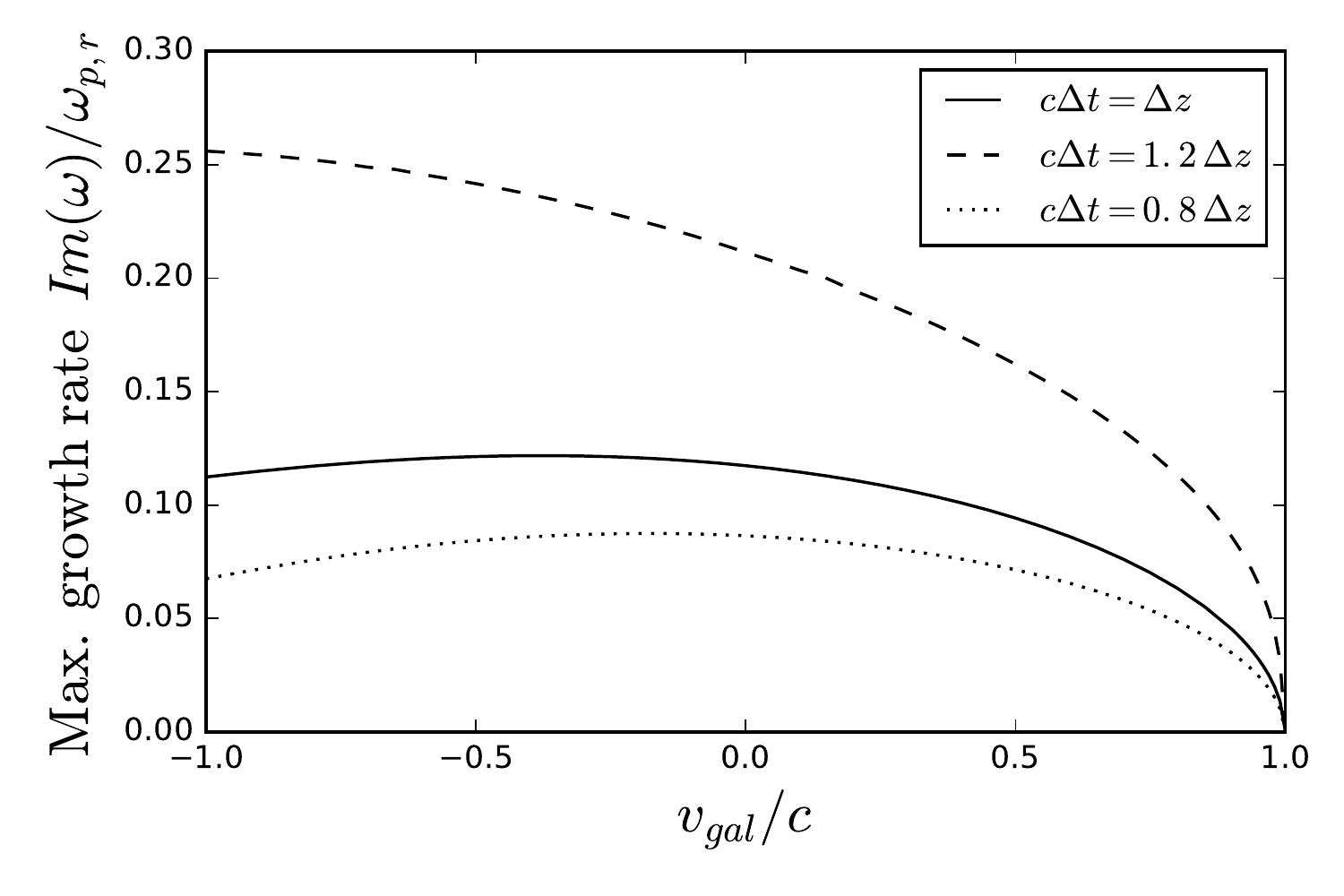}
\caption{\label{fig:scan_vgal}Maximum growth rate of the NCI 
across all spectral modes (i.e. across all $k$
  space), as a function of the velocity $v_{gal}$, and for different
  timesteps $\Delta t$. The growth rates were calculated by solving
  \cref{eq:dispersion} with the parameters from \cref{tab:parameters}.}
\end{figure}

An interesting question is whether $v_{gal}=v_0$ is indeed the optimal
value of the Galilean scheme. To answer this question, we solved the
dispersion equation \cref{eq:dispersion} for a range of value of
$v_{gal}$, spanning from $-c$ to $+c$ (still with the parameters from
\cref{tab:parameters}). In addition, in order to evaluate the robustness of our
scheme with respect to the timestep $\Delta t$, we repeated this
procedure for different values of $\Delta t$. The corresponding growth rates are
plotted in \cref{fig:scan_vgal}, as a function of $v_{gal}$. 
As shown on this figure, the growth of rate of the
instability only goes to 0 for $v_{gal} \simeq c$, thereby confirming
that the optimal Galilean scheme has $v_{gal} \simeq v_0$ (in the case
of an ultrarelativistic plasma). Remarkably, this behavior is observed
for all the values of $\Delta t$ that were tested, thereby indicating
that the Galilean scheme eliminates the NCI independently of the
value of $\Delta t$.

Another important feature of \cref{fig:scan_vgal} is that the growth
rate does not go to 0 for $v_{gal}=-c$. In other words, the NCI is not
suppressed when the relativistic plasma and
Galilean grid move with \emph{opposite} velocities. While this fact is to
be expected from the intuitive picture of \cref{sec:intuitive}, it can potentially
have important implications for practical simulations. For instance, in Lorentz-boosted
simulations of laser-wakefield acceleration, the optimal Galilean
scheme would be that which follows the relativistically-flowing
background plasma (in typical conventions, this plasma flows to the
left). However, in this case, the accelerated electron beam (which
typically moves to the right) would counter-propagate with respect to
the Galilean grid -- thereby potentially triggering the NCI. Nevertheless,
in this particular case, we see no evidence of the NCI in practical
simulations (see \cite{Kirchen2016}), including when analyzing the
emittance of the accelerated beam. This absence of NCI is probably
related to the lack of charge neutrality and limited spatial extent of
the beam, and will be investigated further in the future. In this
regard, one important effect is the fact that the NCI modes often have a group
velocity that is lower than $c$, and thus they rapidly slip behind the beam and
stop growing.

\section{\label{sec:cylindrical}The Galilean PSATD scheme in quasi-cylindrical geometry}

In \cref{sec:cartesian} and \cref{sec:stability-analysis}, we
discussed the Galilean scheme in the context of a spectral \emph{Cartesian}
PIC code. Recently, two spectral \emph{quasi-cylindrical} PIC codes were 
developed \cite{LeheCPC2016,AndriyashPoP2016}. As shown in \cite{Lifschitz},
simulations of physical systems with close-to-cylindrical symmetry can be
made faster by orders of magnitude, when using a quasi-cylindrical
grid instead of a 3D Cartesian grid. Therefore, in the present
section, we extend the Galilean PSATD scheme to the spectral
quasi-cylindrical framework of \cite{LeheCPC2016}.

\subsection{Numerical scheme in quasi-cylindrical geometry}

It was shown in \cite{LeheCPC2016} that a PSATD algorithm could be
derived in quasi-cylindrical geometry, by expressing any scalar field
$S(\vec{x})$ as a sum of Fourier-Bessel modes:
\begin{equation}
S(\vec{x}) \! =\!\!\!\!\!\!\sum_{m=-\infty}^{\infty} \!\!\Integ{k_z} \!\!
\RInteg{k_\perp } \mc{S}_{m}(k_z,k_\perp ) J_m(k_\perp r) \, e^{ik_z z-im\theta} 
\label{eq:CircBwTrans}
\end{equation}
and similarly by expressing any vector field $\vec{V}(\vec{x})$ as
\begin{subequations}
\begin{align}
V_r(\vec{x}) =& \!\!\sum_{m=-\infty}^{\infty} \! 
                \Integ{k_z}\!\!\RInteg{k_\perp }
\left( \mc{V}_{+,m}(k_z,k_\perp ) J_{m+1}(k_\perp r) \right. \nonumber \\
&\left. +\mc{V}_{-,m}(k_z,k_\perp ) J_{m-1}(k_\perp r) \right)  e^{ik_z z-im\theta}
\label{eq:CircBwTransr} 
\end{align}
\begin{align}
V_\theta(\vec{x}) =& \!\!\sum_{m=-\infty}^{\infty} \!
                    \Integ{k_z}\!\!\RInteg{k_\perp }
i\left( \mc{V}_{+,m}(k_z,k_\perp ) J_{m+1}(k_\perp r) \right. \nonumber \\
&\left. - \mc{V}_{-,m}(k_z,k_\perp ) J_{m-1}(k_\perp r) \right)  e^{ik_z z -im\theta } 
\label{eq:CircBwTranst}
\end{align}
\begin{align}
 V_z(\vec{x}) =& \!\!\sum_{m=-\infty}^{\infty} \!\Integ{k_z} \!\!
\RInteg{k_\perp } \times \nonumber \\
& \quad \mc{V}_{z,m}(k_z,k_\perp ) J_m(k_\perp r) \, e^{ik_z z-im\theta} 
\label{eq:CircBwTransz}
\end{align}
\end{subequations}
where $r$ is the radial coordinate, $J_m$ is the Bessel function of
order $m$, and where the sum over $m$ is a sum over azimuthal
modes. (In a practical PIC simulation, this sum is truncated to a low
number of modes, depending on the degree of cylindrical symmetry of the
physical problem.) The terms $\mc{V}_{+,m}$,
$\mc{V}_{-,m}$, $\mc{V}_{z,m}$ and $\mc{S}_m$ represent the spectral
components of the fields $\vec{V}(\vec{x})$ and $S(\vec{x})$.

Within this formalism, the equations of the \emph{standard quasi-cylindrical} PSATD are
very similar to those of the \emph{standard Cartesian} PSATD. In fact,
although the \emph{quasi-cylindrical} equations were derived from first principle in
\cite{LeheCPC2016}, they can alternatively be obtained by using a
formal analogy (see \cref{tab:analogy}) between the representation of the
differential operators in a spectral Cartesian and
spectral quasi-cylindrical framework. More precisely, starting from the
equations of the \emph{standard Cartesian} PSATD, one can obtain the
\emph{standard quasi-cylindrical} PSATD scheme by simply replacing the
expressions in the second line of \cref{tab:analogy} by those in the
third line. 

Therefore here, using the same heuristic procedure, we obtain the equations of
the \emph{Galilean quasi-cylindrical} PSATD (see appendix
\ref{sec:app-cylindricalPSATD} for their full expression) from the equations of the
\emph{Galilean Cartesian} PSATD
\cref{eq:continuity,eq:disc-maxwell1,eq:disc-maxwell2}, by simply
replacing the expressions of the differential operators. Note that, 
in this context, both $\vgal$ and the velocity of the relativistic plasma
$\vec{v}_0$ are necessarily along $z$.

\begin{table*}
\caption{\label{tab:analogy}Representation of common differential
  operators in a spectral Cartesian framework and spectral quasi-cylindrical
  framework (\cref{eq:CircBwTrans,eq:CircBwTransr,eq:CircBwTranst,eq:CircBwTransz}). The
expression of the spectral quasi-cylindrical representation can be derived
by performing the same type of calculation as in appendix B of \cite{LeheCPC2016}.}
\begin{ruledtabular}
\begin{tabular}{c c c c}
Operator
& Gradient: $\vec{F} = \vec{\nabla}S$ & 
Curl: $\vec{F} = \vec{\nabla}\times\vec{V}$ & Divergence: $F = \vec{\nabla}\cdot\vec{V}$\\
\hline
$\begin{array}{c} \mathrm{Spectral\; Cartesian}\\ \mathrm{representation}
\end{array}$ & 
$\begin{array}{c} \vec{\mc{F}} = i\vec{k}\mc{S}\\ i.e. \;
\left\{\begin{array}{l} 
\mc{F}_x = ik_x\mc{S}\\ \mc{F}_y = ik_y\mc{S} \\ \mc{F}_z = ik_z\mc{S} 
\end{array} \right. \end{array}$ &  
$\begin{array}{c} \vec{\mc{F}} = i\vec{k}\times\vec{\mc{V}} \\ i.e.\;
\left\{\begin{array}{l} 
\mc{F}_x = ik_y\mc{V}_z-ik_z\mc{V}_y\\ \mc{F}_y = ik_z\mc{V}_x -
         ik_x\mc{V}_z \\ \mc{F}_z = ik_x\mc{V}_y - ik_y\mc{V}_x
\end{array} \right. \end{array}$ &
$\begin{array}{c} \mc{F} = i\vec{k}\cdot\vec{\mc{V}} \\ i.e. \; 
\mc{F} = ik_x\mc{V}_x + ik_y\mc{V}_y + ik_z\mc{V}_z\end{array}$
\\
\hline 
$\begin{array}{c} \mathrm{Spectral\; cylindrical}\\ \mathrm{representation}
\end{array}$& $\left\{\begin{array}{l} 
\mc{F}_{+,m} = -k_\perp\mc{S}_m/2\\ \mc{F}_{-,m} = k_\perp\mc{S}_m/2 \\ \mc{F}_{z,m} = ik_z\mc{S}_{m}
\end{array} \right.$ &  
$\left\{\begin{array}{l} 
\mc{F}_{+,m} = k_z\mc{V}_{+,m} - ik_\perp\mc{V}_{z,m}/2 \\ 
\mc{F}_{-,m} =  - k_z\mc{V}_{-,m} -ik_\perp\mc{V}_{z,m}/2 \\ 
\mc{F}_{z,m} = ik_\perp\mc{V}_{+,m} + ik_\perp \mc{V}_{-,m}
\end{array} \right.$ & 
$\mc{F}_m = k_\perp(\mc{V}_{+,m} - \mc{V}_{-,m}) + ik_z\mc{V}_{z,m}$\\
\end{tabular}
\end{ruledtabular}
\end{table*}

Apart from these modified equations, the structure of our PIC cycle in
quasi-cylindrical geometry is identical to that presented in \cref{sec:scheme-cartesian} for
Cartesian geometry.

\subsection{Stability of a uniform, relativistic plasma}

The Galilean PSATD scheme described in the previous section was
implemented in the spectral quasi-cylindrical code \textsc{FBPIC}
\cite{LeheCPC2016}. We then performed test simulations featuring a
uniform relativistic plasma. Apart from the shape factor (which was
set to order 1), the numerical and physical parameters of the
simulations are the same as in \cref{tab:parameters} (where $\Delta
x$ and $N_x$ are to be replaced with the corresponding radial
parameters $\Delta r=0.3868 \,k_{p,r}^{-1}$
and $N_r = 100$). In addition, the spatial smoothing function $\mc{T}$
was set to $\mc{T}(k_z, k_\perp) = \cos^2(k_z\Delta z/2)\cos^2(k_\perp
\Delta r/2)$ as in \cite{LeheCPC2016}.

\begin{figure}
\includegraphics[width=\columnwidth]{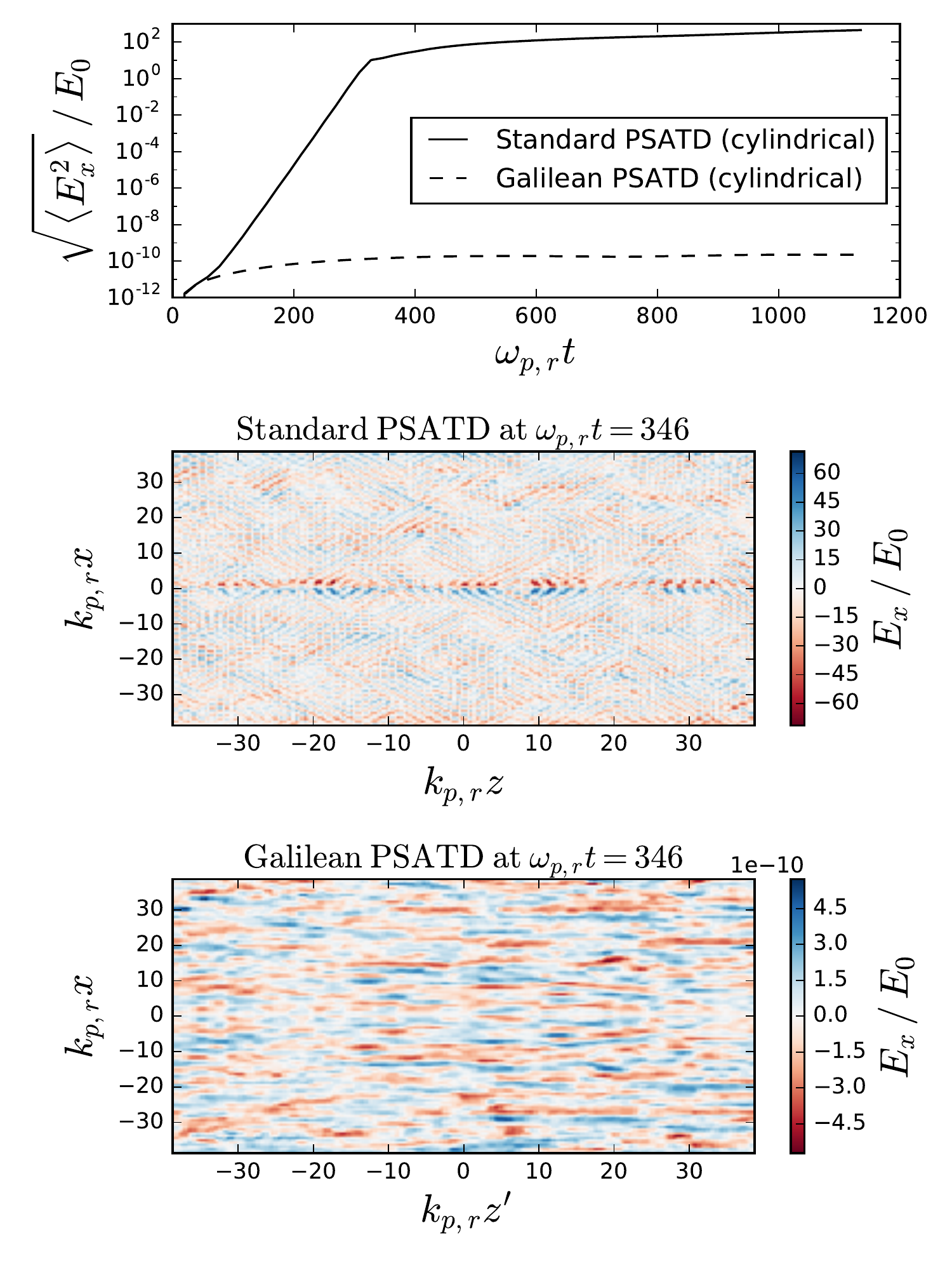}
\caption{\label{fig:empirical-circ}Results of a relativistic flowing
  plasma simulation in the quasi-cylindrical PIC code FBPIC -- with
  both the standard and Galilean PSATD scheme. Top panel: evolution of
  the RMS amplitude of the electric field in the simulation box. Middle and
  bottom panels: Maps of the electric field in the simulation box, at a given time.}
\end{figure}

The results of these simulations are represented in
\cref{fig:empirical-circ}, using a similar layout as for the
corresponding Cartesian simulation (see \cref{fig:empirical}). These
quasi-cylindrical simulations support the same conclusions as 
their Cartesian counterpart: the standard PSATD scheme is unstable
due to the NCI (as evidenced by the solid line in the upper panel of
\cref{fig:empirical-circ} and by the high-frequency pattern in the
corresponding field map, on the middle panel), while the Galilean
PSATD scheme remains stable (see the dashed line in the upper panel,
and the corresponding field map on the bottom panel, which are
consistent with numerical and thermal noise).

\section*{Conclusion and discussion}

In this article, we showed that integrating the PIC equations in
Galilean coordinates supresses the NCI, for a plasma drifting at a uniform relativistic velocity -- both in Cartesian and
quasi-cylindrical geometry. This new numerical scheme opens promising
possibilities, especially for Lorentz-boosted simulations of
laser-wakefield acceleration -- as shown in \cite{Kirchen2016}.

Since the supression of the NCI is the aim of a number of previous
schemes \cite{GodfreyJCP2014,GodfreyIEEE2014,GodfreyJCP2014b,
Godfreyarxiv2014,GodfreyCPC2015,YuCPC2015,YuCPC2015-Circ,Yu-arxiv2016}, 
it is worth discussing here the advantages and drawbacks of the 
Galilean PSATD scheme in relation to previous work, as well as areas
of possible improvements.

As mentioned in the introduction, one advantage of the Galilean scheme
is that it is built on the natural integration of the Maxwell equations, and
does not introduce strong smoothing, or arbitrary or manually-tuned numerical
corrections. This contrasts for instance with 
\cite{GodfreyJCP2014,GodfreyIEEE2014,GodfreyJCP2014b,Godfreyarxiv2014}, 
but also with the methods from
\cite{YuCPC2015,YuCPC2015-Circ,Yu-arxiv2016} 
in which both the timestep and ``bump'' in $k$ space need to be tuned in relation with
the plasma density \cite{YuCPC2015} (making it potentially difficult 
to simulate plasmas with longitudinally or transversally varying
density profiles). On the other hand, while the methods from
\cite{GodfreyJCP2014,GodfreyIEEE2014,GodfreyJCP2014b,Godfreyarxiv2014,YuCPC2015,YuCPC2015-Circ,Yu-arxiv2016} 
can in some cases simulate relativistically crossing plasmas, in the present
formulation of the Galilean scheme this could trigger the NCI for one
of the two crossing plasmas (see \cref{sec:influence-vgal}). 
In future works, we will present an alternate formulation of
the Galilean scheme, using multiple grids for the current $\vec{J}$, 
which relaxes the restrictions on simulations of crossing plasmas.

Another important point is that the usual advantages of the standard
PSATD scheme naturally carry over to the Galilean PSATD scheme,
including dispersion-free wave propagation (in all directions) 
and suppression of staggered interpolation artifacts (see
e.g. \cite{LeheCPC2016}). This is not the case for
methods based on Finite-Difference Time-Domain (FDTD), Pseudo-Spectral
Time-Domain (PSTD) or hybrid schemes (PSTD longitudinally and FDTD
transversally), as in \cite{GodfreyJCP2014b,
Godfreyarxiv2014,GodfreyCPC2015,YuCPC2015,YuCPC2015-Circ,Yu-arxiv2016}. 
On the other hand, the methods based on FDTD or hybrid schemes 
can be more easily scaled to multiple computing nodes (see
esp. \cite{Yu-arxiv2016}). In order to mitigate this limitation, the
Galilean scheme is using domain decomposition, as proposed in 
\cite{VayJCP2013} in both Warp and FBPIC. To remove the limitation
further, this was extended in Warp (and will be extended in FBPIC in
the future) to incorporate a spectral
representation of \emph{finite-difference} high-order operators (as
discussed e.g. in \cite{Birdsall2004appE,VayAAC2014,Vincenti2015,Yu-arxiv2016}), which have better
scalability than purely-spectral operators.

\begin{acknowledgments}
The simulation results were stored and visualized using the new
open-source format openPMD \cite{openPMD}. The authors wish to thank the
openPMD contributors, and in particular its creator Axel Huebl
(HZDR, Germany). The authors also thank Patrick Lee (U. Paris-Sud,
France) for interesting discussions and for performing additional
tests of the Galilean PSATD scheme (not presented here).

This work was partly supported by the Director, Office of Science, Office of High Energy Physics, U.S. Dept. of Energy under Contract No. DE-AC02-05CH11231, including from the Laboratory Directed Research and Development (LDRD) funding from Berkeley Lab.

This document was prepared as an account of work sponsored in part by the United States Government. While this document is believed to contain correct information, neither the United States Government nor any agency thereof, nor The Regents of the University of California, nor any of their employees, nor the authors makes any warranty, express or implied, or assumes any legal responsibility for the accuracy, completeness, or usefulness of any information, apparatus, product, or process disclosed, or represents that its use would not infringe privately owned rights. Reference herein to any specific commercial product, process, or service by its trade name, trademark, manufacturer, or otherwise, does not necessarily constitute or imply its endorsement, recommendation, or favoring by the United States Government or any agency thereof, or The Regents of the University of California. The views and opinions of authors expressed herein do not necessarily state or reflect those of the United States Government or any agency thereof or The Regents of the University of California.
\end{acknowledgments}

\appendix
\section{\label{app:Maxwell-integration}Analytical integration of the Maxwell equations from $t=n\Delta t$ to $t=(n+1)\Delta t$}

After inserting \cref{eq:J-evolution,eq:rho-evolution}
into \cref{eq:spectral-2ndorder1,eq:spectral-2ndorder2}, we obtain
the following equations:
\begin{subequations}
\begin{equation}
\left( \Dt{\;} -i\vec{k}\cdot\vgal\right)^2 \vec{\mc{B}} + c^2\vec{k}^2 \vec{\mc{B}}
 =  \frac{1}{\epsilon_0}i\vec{k}\times\vec{\mc{J}}^{n+1/2} \label{eq:app-2ndorder1}
\end{equation}
\begin{align}
\left( \Dt{\;} - i\vec{k}\cdot\vgal\right)^2 &\vec{\mc{E}} + c^2\vec{k}^2 \vec{\mc{E}}
 =  \frac{1}{\epsilon_0}i(\vec{k}\cdot\vgal)\vec{\mc{J}}^{n+1/2} \nonumber\\
& -\frac{c^2}{\epsilon_0}\mc{\rho}^{n+1} \frac{1 -
  e^{i\vec{k}\cdot\vgal(t-n\Delta t)}}{1-e^{i\vec{k}\cdot\vgal\Delta t}} \,i\vec{k} \nonumber \\ 
& +\frac{c^2}{\epsilon_0}\mc{\rho}^n \frac{e^{i\vec{k}\cdot\vgal\Delta t} -
  e^{i\vec{k}\cdot\vgal(t-n\Delta t)}}{1-e^{i\vec{k}\cdot\vgal\Delta t}} \,i\vec{k}\label{eq:app-2ndorder2}
\end{align}
\end{subequations}
where $\mc{\rho}^{n+1}$, $\mc{\rho}^n$ and $\vec{\mc{J}}^{n+1/2}$ are
constant. Notice in particular that the time derivative of
$\vec{\mc{J}}$ from \cref{eq:spectral-2ndorder2} vanishes in
\cref{eq:app-2ndorder2} due to \cref{eq:J-evolution}.

For the purpose of time integration, both equations can be cast into the
following general form:
\begin{equation}
\label{eq:app-general-equadiff}
\left( \Dt{\;} -i\nu c k\right)^2 f + c^2k^2 f
 = \alpha + \beta e^{i\nu c k (t-n\Delta t)}
\end{equation}
where $\nu \equiv
\vec{k}\cdot\vgal /ck$ as in \cref{eq:def-nu-theta}, and where $\alpha$ and $\beta$ are constants. For instance, in the case of
\cref{eq:app-2ndorder1}, one has:
\begin{equation}
\label{eq:app-alpha-beta1}
\alpha = \frac{1}{\epsilon_0}i\vec{k}\times\vec{\mc{J}}^{n+1/2} \qquad
\beta = 0 \qquad f = \vec{\mc{B}}
\end{equation}
while in the case of \cref{eq:app-2ndorder2}:
\begin{align}
\label{eq:app-alpha-beta2}
& \alpha = \frac{i \nu c k}{\epsilon_0}\vec{\mc{J}}^{n+1/2}  -
\frac{c^2}{\epsilon_0} \frac{\mc{\rho}^{n+1} -
  \mc{\rho}^ne^{i\nu c k\Delta t}
  }{1-e^{i\nu c k\Delta t}} i\vec{k} \nonumber \\
& \beta = \frac{c^2}{\epsilon_0} \frac{\mc{\rho}^{n+1} -
  \mc{\rho}^n }{1-e^{i\nu c k\Delta t}} i\vec{k} \qquad f = \vec{\mc{E}}
\end{align}
The general solution of \cref{eq:app-general-equadiff} is:
\begin{align}
\label{eq:app-general-solution}
f(t) =& \kappa_1 \cos(ck(t-n\Delta t) ) e^{i\nu ck(t-n\Delta t)}
        \nonumber \\
& + \kappa_2 \sin(ck(t-n\Delta t) ) e^{i\nu ck(t-n\Delta t) } \nonumber \\
&+ \frac{\alpha}{c^2k^2(1-\nu^2)} + \frac{\beta
  }{c^2k^2}e^{i\nu c k (t-n\Delta t)}
\end{align}
where $\kappa_1$ and $\kappa_2$ are integration constants. These
integration constants can be
determined from the initial condition $f(n\Delta t)$ and
$\partial_tf(n\Delta t)$, in which case \cref{eq:app-general-solution}
becomes:
\begin{widetext}
\begin{align}
f(t) =& \left[ f(n\Delta t) - \frac{\alpha}{c^2k^2 (1-\nu^2)} - \frac{\beta}{c^2k^2}  \right]\cos(ck(t-n\Delta t) ) e^{i\nu c k(t-n\Delta t)} + \frac{\alpha}{c^2k^2 (1-\nu^2)} + \frac{\beta
  }{c^2k^2}e^{i\nu c k (t-n\Delta t)}\nonumber \\
&+ \frac{1}{ck}\left[\partial_t f(n\Delta t) - i\nu c k f(n\Delta t) +
  i\nu c k \frac{\alpha}{c^2k^2 (1-\nu^2)} \right] \sin(ck(t-n\Delta
  t))e^{i\nu c k(t-n\Delta t) } 
\end{align}
\end{widetext}

Finally, since our purpose is to integrate
\cref{eq:app-2ndorder1,eq:app-2ndorder2} from $t=n\Delta t$ to
$t=(n+1)\Delta t$, let us evaluate the above equation at
$t=(n+1)\Delta t$:
\begin{align}
\label{eq:app-disc-solution}
f((n+1)\Delta t) &= C\theta^2 f(n\Delta t) + \frac{\theta\chi_1}{c^2k^2}
                   \alpha + \frac{\theta^2 (1-C) }{c^2k^2}\beta \nonumber \\
&+ \frac{S\theta^2}{ck} [ \partial_tf(n\Delta t) - i\nu c k f(n\Delta t) ]  
\end{align}
where $C$, $S$, $\theta$ and $\chi_1$ have the same definition as in
\cref{eq:def-C-S,eq:def-nu-theta,eq:def-chi1}. The integrated Maxwell
equations \cref{eq:disc-maxwell1,eq:disc-maxwell2} are then obtained by combining
\cref{eq:app-disc-solution} with
\cref{eq:app-alpha-beta1,eq:app-alpha-beta2} respectively. In
particular, in order to evaluate the last term in
\cref{eq:app-disc-solution}, we used the equations:
\begin{subequations}
\begin{align}
\left( \Dt{\;} - i\nu ck \right) \vec{\mc{B}} = - i\vec{k}\times
  \vec{\mc{E}} \\
\left( \Dt{\;} - i\nu ck \right) \vec{\mc{E}} = c^2 i\vec{k}\times
  \vec{\mc{B}} - \frac{1}{\epsilon_0}\vec{\mc{J}}
\end{align}
\end{subequations}
which are the spectral representations of the Maxwell equations \cref{eq:maxwell1,eq:maxwell2}.

\section{\label{sec:derivation-dispersion}Derivation of the dispersion relation, for the Galilean PSATD
scheme}

The dispersion relation typically results from combining the Vlasov equation and Maxwell
equations. Here, we use a discretized version of the Vlasov equation
and Maxwell equation that take into account all the numerical effects
described in section \cref{sec:scheme-cartesian} (interpolation to
grid; current correction, etc.).

We consider a periodic box, and a uniform plasma having a density $n_0$
and a relativistic factor $\gamma_0$. We will treat perturbations
$\delta f$ to the distribution function, as well as the fields
$\vec{E}$ and $\vec{B}$, as small quantities. 

\subsection{Notations and definitions}

Let us consider a 2D Cartesian grid with $N_x \times N_z$
gridpoints and periodic boundaries. We will denote the position of the
gridpoints $\xj$, i.e. 
\begin{equation}
\label{eq:app-def-xj}
\xj = j_x \Delta x \,\vec{u}_x + j_z \Delta z \,\vec{u}_z \quad 
\begin{array}{c}
j_x \in [0, N_x-1] \\
j_z \in [0, N_z -1]
\end{array}
\end{equation}
where $j_x$ and $j_z$ are integers. In addition, we will denote $\Xll$
the vectors of periodicity of the grid, i.e. 
\begin{equation}
\label{eq:app-def-Xll}
\Xll = \ell_x N_x \Delta x \,\vec{u}_x + \ell_z N_z\Delta z \,\vec{u}_z \qquad
\ell_x \in \mathbb{Z}, \ell_z \in \mathbb{Z}
\end{equation}
With these notations, any vector of the reciprocal lattice can be
written as $\km = \vec{k} + \Km$ where $\vec{k}$ is a vector of the first
Brillouin zone, and $\Km$ is a vector of periodicity of the reciprocal
lattice i.e. $\vec{k}$ and $\Km$ are of the form:
\begin{align}
&\Km = m_x \frac{2\pi}{\Delta x}\vec{u}_x + m_z \frac{2\pi}{\Delta
  z}\vec{u}_z \qquad
m_x \in \mathbb{Z}, m_z \in \mathbb{Z} \label{eq:app-def-k}\\
&\vec{k} = j_x \frac{2\pi}{N_x\Delta x} \vec{u}_x +
j_z\frac{2\pi}{N_z\Delta z}\vec{u}_z \quad 
\begin{array}{c}
j_x \in [-N_x/2, N_x/2-1] \\
j_z \in [-N_z/2, N_z/2 -1]
\end{array}
\label{eq:app-def-Km}\end{align}
where $j_x$ and $j_z$ are integers.

With these definitions, the expressions of the discrete Fourier
transforms of the grid fields $\vec{E}$ and
$\vec{B}$ (which are defined \emph{exclusively} on the gridpoints $\xj$) are:
\begin{equation}
\label{eq:app-def-FourierEB}
\vec{\mc{E}}(\vec{k}) = \frac{\sum_{\vec{j}}
  e^{-i\vec{k}\cdot\xj}\vec{E}(\xj)}{N_xN_z} \qquad \vec{\mc{B}}(\vec{k}) = \frac{\sum_{\vec{j}}
  e^{-i\vec{k}\cdot\xj}\vec{B}(\xj)}{N_xN_z} 
\end{equation}
By contrast, the expression of the Fourier transform for the distribution function
$f(\vec{x}',\vec{p})$ (which is defined also \emph{inbetween}
gridpoints, and is periodic) is:
\begin{equation}
\label{eq:app-def-Fourierf}
\hat{f}(\vec{k} + \Km,\vec{p}) = \frac{1}{N_xN_z \Delta x \, \Delta
  z}\int_{box} \!\!\!\!\!\!\!\! d\vec{x}'\,  e^{-i(\vec{k}+\Km)\cdot\vec{x'}} f(\vec{x'},\vec{p})
\end{equation}
where the integration is performed over the (finite) extent of the
box.

Finally, the Fourier transform of the particle shape factor $S$ (which
is defined over $\mathbb{R}^2$ and is not periodic -- since for
instance $S(\vec{x})=(1 - |x|/\Delta x)\Theta(\Delta x - |x|)\times
(1 - |z|/\Delta z)\Theta(\Delta z - |z|)$ for order 1 shape factor)
is:
\begin{equation}
\label{eq:app-def-FourierS}
\mc{S}(\vec{k} + \Km) = \frac{1}{\Delta x\Delta z}\int_{\mathbb{R}^2}d\vec{x}' \, e^{-i\vec{k} + \Km\cdot\vec{x}'}S(\vec{x}')
\end{equation}
These distinctions regarding the Fourier transform are important in order to correctly derive the
space aliases.

\subsection{Discretized Vlasov equation}

Let us define $f^{n+1/2}(\vec{x}',\vec{p})$ as the distribution function of positions and
momenta at half-integer step $(n+1/2)\Delta t$, and let us derive the
evolution of the $f$ from one half-integer step to the next.

From the equations of motion of the particles
\cref{eq:motion2,eq:disc-motion}, the evolution of position and
momenta of one given particle from one half-integer timestep to the
next is:
\begin{subequations}
\begin{align}
\frac{\vec{x}'^{n+1/2} - \vec{x}'^{n-1/2}}{\Delta t} &=
                                                        \frac{\vec{p}^{n+1/2}}{2m\gamma^{n+1/2}} + \frac{\vec{p}^{n-1/2}}{2m\gamma^{n-1/2}} - \vgal \\
\frac{\vec{p}^{n+1/2}-\vec{p}^{n-1/2}}{\Delta t} &=
  q \vec{E}^n_{interp} +  \nonumber\\
& q\left(\frac{\vec{p}^{n+1/2}}{2m\gamma^{n+1/2}} + \frac{\vec{p}^{n-1/2}}{2m\gamma^{n-1/2}}\right)\times\vec{B}^n_{interp}
\end{align}
\end{subequations}
where by definition $\vec{x}'^{n+1/2} \equiv
(\vec{x}'^{n+1}+\vec{x}'^n)/2$, and where $\vec{E}^n_{interp}$ and
$\vec{B}^n_{interp}$ are the interpolated fields at the particle's
position, at time $n\Delta t$.

Since the fields $\vec{E}$ and $\vec{B}$ are treated as small quantities, the modifications of $\vec{p}$ over one timestep is small, and thus these equations can be approximated to:
\begin{subequations}
\begin{align}
\frac{\vec{x}'^{n+1/2} - \vec{x}'^{n-1/2}}{\Delta t} =&
\frac{\vec{p}^{n-1/2}}{m\gamma^{n-1/2}} - \vgal \\
\frac{\vec{p}^{n+1/2}-\vec{p}^{n-1/2}}{\Delta t} =&
  q \vec{E}^n_{interp} + q\frac{\vec{p}^{n-1/2}}{m\gamma^{n-1/2}}\times\vec{B}^n_{interp}
\end{align}
\end{subequations}

Because the volume in phase space is conserved during this
evolution, the corresponding evolution of the distribution function
$f$ is:
\begin{align}
&f^{n+1/2}\left[ \vec{x}' +
   \left(\frac{\vec{p}}{m\gamma} -
   \vgal\right)\Delta t, \right. \nonumber \\
 &\left. \vec{p} + q \Delta t\left(\vec{E}^n_{interp}+ \frac{\vec{p}}{m\gamma}\times\vec{B}^n_{interp}\right)\right]
   =  f^{n-1/2}(\vec{x}', \vec{p})
\end{align}
Now $f^{n+1/2}(\vec{x}',\vec{p})$ is of the form $f_0(\vec{p}) +
\delta f^{n+1/2}(\vec{x}',\vec{p})$ where $f_0$ is the distribution
function of a uniform, stationary plasma and $\delta f$ is a perturbation.
Since $\delta f$, $\vec{E}$ and $\vec{B}$ are treated as small
perturbations, the above equation can be Taylor-expanded to first order:
\begin{align} 
\label{eq:app-pre-vlasov}
& \delta f^{n+1/2}(\vec{x}' + \vec{v}\Delta t - \vgal\Delta t, \vec{p}) - \delta f^{n-1/2}(\vec{x}', \vec{p})  \nonumber \\
& \qquad + q\Delta t\left( \vec{E}^n_{interp} + \vec{v}\times
  \vec{B}^n_{interp}\right)\cdot \frac{\partial f_0}{\partial \vec{p}} = 0
\end{align}
where we used the short-hand notation $\vec{v} \equiv \vec{p}/(\gamma m)$.

Now since $\vec{E}^{n}_{interp}$ is interpolated to the macroparticles at
time $n\Delta t$, its expression for a
given macroparticle at position $\vec{x}'^n$ is:
\begin{align}
\vec{E}^n_{interp} &= \sum_{\vec{j},\vec{\ell}}
S(\vec{x}'^n-\xj - \Xll) \vec{E}^n(\xj + \Xll) \nonumber \\
&= \sum_{\vec{j},\vec{\ell}}
S(\vec{x}'^n-\xj - \Xll) \vec{E}^n(\xj ) 
\label{eq:app-interp-E}
\end{align}
where $\Xll$ are vectors of periodicity of the
grid and $\xj$ denote gridpoints (see
\cref{eq:app-def-xj,eq:app-def-Xll}) and 
where the sum over $\vec{j}$ corresponds to a sum over the whole
(finite) grid. $S$ is the shape factor of the macroparticle. Finally $\vec{E}^n(\xj)$ is the expression of
$\vec{E}$ on the grid. Since $\vec{B}$ is defined on the same grid and
at the same time as $\vec{E}$ (i.e. the grid is not staggered here),
the equation for $\vec{B}^n_{interp}$ is similar to
\cref{eq:app-interp-E}. By combining \cref{eq:app-interp-E} with
\cref{eq:app-pre-vlasov} and the equation $\vec{x}'^{n} =
\vec{x}'^{n-1/2} + [ \; \vec{p}^{n-1/2}/(\gamma^{n-1/2} m) -
\vgal\;]\Delta t/2$, we obtain:
\begin{align} 
& \delta f^{n+1/2}(\vec{x}' + \vec{v}\Delta t - \vgal\Delta t, \vec{p}) - \delta f^{n-1/2}(\vec{x}', \vec{p}) + \nonumber \\
&  \qquad + q\Delta t \sum_{\vec{j},\vec{\ell}}S\left(\vec{x}' +
  \frac{(\vec{v}-\vec{v}_{gal})\Delta t}{2} -\xj - \Xll \right) \nonumber\\
 &\qquad \qquad \times \left( \vec{E}^n(\xj) + \vec{v}\times \vec{B}^n(\xj)\right)\cdot \frac{\partial f_0}{\partial \vec{p}} = 0
\end{align}

Let us evaluate the Fourier transform of the above equation, at a vector
of the reciprocal lattice $\vec{k}_m = \vec{k} +\Km$ (see
\cref{eq:app-def-k,eq:app-def-Km}). By using the definition
\cref{eq:app-def-Fourierf}, we have:
\begin{align}
&\delta \hat{f}^{n+1/2}(\km,\vec{p})
  \,e^{i\km\cdot(\vec{v}-\vec{v}_{gal})\Delta t} - \delta
  \hat{f}^{n-1/2}(\km,\vec{p}) \nonumber \\
& + \frac{q\Delta t}{\Delta x \, \Delta z} \sum_{\vec{j}} \left[ \int_{\mathbb{R}^2} d\vec{x}'
  \,e^{-i\km\cdot\vec{x}'}S\left(\vec{x}' + \frac{(\vec{v}-\vec{v}_{gal})\Delta t}{2} -
  \xj \right)\right] \nonumber \\ 
&\qquad \times \frac{1}{N_xN_z}\left( \vec{E}^n(\xj) + \vec{v}\times \vec{B}^n(\xj)
  \right)\cdot \frac{\partial f_0}{\partial \vec{p}} = 0 
\end{align}
And finally, by using \cref{eq:app-def-FourierEB,eq:app-def-FourierS}:
\begin{align}
\delta &\hat{f}^{n+1/2}(\km,\vec{p})
  \,e^{i\km\cdot(\vec{v}-\vec{v}_{gal})\Delta t/2} \nonumber \\
&- \delta
  \hat{f}^{n-1/2}(\km,\vec{p})\,e^{-i\km\cdot(\vec{v}-\vec{v}_{gal})\Delta
  t/2} \nonumber \\
& + q \Delta t\,\mc{S}(\km) \left( \vec{\mc{E}}^n(\vec{k}) + \vec{v}\times \vec{\mc{B}}^n(\vec{k})
  \right)\cdot \frac{\partial f_0}{\partial \vec{p}} = 0 
\label{eq:app-disc-vlasov}
\end{align}

\subsection{Discretized Maxwell equation}

Let us now derive an expression of the discretized Maxwell equations,
where the source terms are expressed as a function of $\delta f$. Let
us first remark that, in \cref{eq:disc-maxwell1,eq:disc-maxwell2}, the
terms $\mc{\rho}^n$, $\mc{\rho}^{n+1}$ and
$\vec{\mc{J}}^{n+1/2}$ are the charge and current obtained
\emph{after} current correction and smoothing, as described in
\cref{sec:scheme-cartesian}. After inserting the
explicit expression for current correction and smoothing
\cref{eq:current-correction,eq:smoothing}, the discretized Maxwell
equations become:
\begin{subequations}
\begin{equation}
\vec{\mc{B}}^{n+1} = \theta^2 C \vec{\mc{B}}^n
 -\frac{\theta^2 S}{ck}i\vec{k}\times \vec{\mc{E}}^n
+ \;\frac{\theta \chi_1 \mc{T}}{\epsilon_0c^2k^2}\;i\vec{k} \times
                     \vec{\mc{J}}_d^{n+1/2} \label{eq:app-nonsym-maxwell1}
\end{equation}
\begin{align}
\vec{\mc{E}}^{n+1} &=  \theta^2 C  \vec{\mc{E}}^n
 +\frac{\theta^2 S}{k} \,c i\vec{k}\times \vec{\mc{B}}^n - \frac{\mc{T} i\vec{k} }{\epsilon_0k^2}\left(\mc{\rho}^{n+1} -
  \theta^2C\mc{\rho}^{n} \;\right)\nonumber \\
&+\frac{i\nu \theta \chi_1 - \theta^2S}{\epsilon_0 ck} \; \mc{T}\left( \vec{\mc{J}}^{n+1/2}_d -
  \frac{(\vec{k}\cdot\vec{\mc{J}}^{n+1/2}_d)\vec{k}}{k^2}  \right) \label{eq:app-nonsym-maxwell2}
\end{align}
\end{subequations}
where $\mc{\rho}^n$, $\mc{\rho}^{n+1}$ and
$\vec{\mc{J}}_d^{n+1/2}$ are the charge and current obtained just
after deposition and \emph{before} current correction and smoothing, and where $\mc{T}$ is the
smoothing factor defined in \cref{eq:smoothing} (with $k_y=0$ in the
2D Cartesian case).

In addition, the above equations can be rewritten in a time-symmetrical
form, which is more convenient for the analysis in the rest of this
appendix. It can indeed be verified that, if $\vec{\mc{E}}$ and
$\vec{\mc{B}}$ satisfy
\cref{eq:app-nonsym-maxwell1,eq:app-nonsym-maxwell2}, as well as the
associated conservation equations $i\vec{k}\cdot\vec{\mc{B}}^n=0$ and $i\vec{k}\cdot\vec{\mc{E}}^n=\mc{T}\mc{\rho}^n/\epsilon_0$, then they also satisfy
\begin{align}
&\theta^* c\vec{\mc{B}}^{n+1} - \theta c \vec{\mc{B}}^{n} =
-t_{ck}\frac{i\vec{k}\times 
(\theta^*\vec{\mc{E}}^{n+1} + \theta \vec{\mc{E}}^n)}{k} \nonumber \\
& \qquad + 2\chi_4'\frac{ \mc{T}}{\epsilon_0 ck}
  \frac{\vec{k}\times\vec{\mc{J}_d}^{n+1/2}}{k} \label{eq:app-disc-maxwell1}
\end{align}
\begin{align}
&\theta^* \vec{\mc{E}}^{n+1} - \theta \vec{\mc{E}}^{n} = t_{ck}\frac{i\vec{k}\times 
(\theta^*c\vec{\mc{B}}^{n+1} + \theta c\vec{\mc{B}}^n)}{k} \nonumber \\
& \qquad  - \frac{\mc{T} i\vec{k} }{\epsilon_0k^2}(
  \theta^*\mc{\rho}^{n+1} - \theta\mc{\rho}^{n}) \nonumber \\
& \qquad - 2\chi_4\frac{\mc{T}}{\epsilon_0 c k}\left( \vec{\mc{J}}^{n+1/2}_d -
  \frac{(\vec{k}\cdot\vec{\mc{J}}^{n+1/2}_d)\vec{k}}{k^2}  \right) \label{eq:app-disc-maxwell2}
\end{align}

where $^*$ denotes the complex conjugate, and where 
\begin{align}
&\chi_4 
=  \frac{c_{\nu ck}(t_{ck} 
  - \nu t_{\nu ck}) }{(1-\nu^2)} \qquad \chi_4' 
= \frac{c_{\nu ck}(t_{\nu ck}
  - \nu t_{ck}) }{(1-\nu^2)} \\
&t_{ck} \equiv \tan\left(\frac{ck\Delta t}{2}\right) \qquad c_{ck}
  \equiv \cos\left(\frac{ck\Delta t}{2}\right) \\
&t_{\nu ck} \equiv \tan\left(\frac{\nu ck\Delta t}{2}\right) \qquad
  c_{\nu ck} \equiv \cos\left(\frac{\nu ck\Delta t}{2}\right) 
\end{align}

Let us now express the deposited charge and currents as a function of
$\delta f$. The current density, which is deposited at half-integer time, is given within one given cell by
\begin{equation}
\vec{J}_d^{n+\frac{1}{2}}\!(\xj) = \frac{1}{\Delta x \Delta z}\!\!\int \!\!  d\vec{p} \!\!\int_{\mathbb{R}^2}\!\!\!\!\! d\vec{x'}
\;q\vec{v}\, S(\xj-\vec{x}')\delta
f^{n+1/2}(\vec{x'}, \vec{p}) 
\end{equation}
In the above expression, the integration is carried out over all space
($\mathbb{R}^2$) because the shape factor $S$ may extend beyond the
finite grid. Let us now expand the periodic function $\delta f$ in Fourier series:
\begin{align}
\vec{J}_d^{n+\frac{1}{2}}\!(\xj) =& \frac{1}{\Delta x \Delta z}\!\!\int \!\!  d\vec{p} \!\!\int_{\mathbb{R}^2}\!\!\!\!\! d\vec{x'}
\;q\vec{v}\, S(\xj-\vec{x}') \nonumber \\ 
&\times\sum_{\km}\delta \hat{f}^{n+1/2}(\km, \vec{p}) e^{i\km\cdot\vec{x}'}
\end{align}
where the sum is over all vectors $\km = \vec{k}+\Km$ of the reciprocal lattice (see
\cref{eq:app-def-k,eq:app-def-Km}). With some algebra, this can be
rewritten as:
\begin{equation}
\vec{J}_d^{n+\frac{1}{2}}\!(\xj) = \sum_{\vec{k}} e^{i\vec{k}\cdot\xj}\sum_{\vec{m}}
\mc{S}(\km) \!\!\int \!\! d\vec{p} \;q\vec{v}\,\delta \hat{f}^{n+1/2}(\km, \vec{p})
\end{equation}
where we used the relation $e^{i\Km\cdot\xj}=1$, which comes from
\cref{eq:app-def-xj,eq:app-def-Km}. By identification, we have
\begin{equation}
\vec{\mc{J}}_d^{n+\frac{1}{2}}(\vec{k}) = \sum_{\vec{m}}
\mc{S}(\km) \!\!\int \!\! d\vec{p} \;q\vec{v}\,\delta
\hat{f}^{n+1/2}(\km, \vec{p})
\label{eq:app-source-J}
\end{equation}

Similarly, since the charge density $\rho^n$ is deposited from the particle position $\vec{x'}^n = \vec{x'}^{n+1/2} - (\vec{v}^{n+1/2}-\vgal)\Delta t/2$, its expression is:
\begin{align} 
\rho^{n} (\xj) =& \frac{1}{\Delta x \Delta z}\!\!\int \!\!  d\vec{p} \!\!\int_{\mathbb{R}^2}\!\!\!\!\! d\vec{x'}
\;q\, S\left(\xj-\vec{x}' + \frac{(\vec{v}-\vec{v}_{gal})\Delta
  t}{2}\right) \nonumber \\
& \qquad \times \delta f^{n+1/2}(\vec{x'}, \vec{p}) 
\end{align}
And thus the expression of $\mc{\rho}^{n}$ is:
\begin{equation}
\mc{\rho}^{n}(\vec{k}) = \sum_{\vec{m}}
\mc{S}(\km) \!\!\int \!\! d\vec{p} \;q\,\delta \hat{f}^{n+1/2}(\km,
\vec{p})e^{\frac{i\km\cdot(\vec{v}-\vgal)\Delta t}{2}} \label{eq:app-source-rho1}
\end{equation}
And similarly the expression of $\mc{\rho}^{n+1}$ is:
\begin{equation}
\mc{\rho}^{n+1}(\vec{k}) = \sum_{\vec{m}}
\mc{S}(\km) \!\!\int \!\! d\vec{p} \;q\,\delta \hat{f}^{n+1/2}(\km,
\vec{p})e^{-\frac{i\km\cdot(\vec{v}-\vgal)\Delta t}{2}} \label{eq:app-source-rho2}
\end{equation}

\subsection{Eigenmodes and eigensystem}

The discretized Vlasov equation \cref{eq:app-disc-vlasov}, the discretized Maxwell
equations \cref{eq:app-disc-maxwell1,eq:app-disc-maxwell2} and the
expression of the source terms
\cref{eq:app-source-J,eq:app-source-rho1,eq:app-source-rho2} form a
set of coupled equations of evolution. Let us look for eigenmodes of
this set of equations, where we assume all perturbations to be of the
form $e^{i\vec{k}\cdot\vec{x}-i\omega
  t}= e^{i\vec{k}\cdot\vec{x}' -i(\omega-\vec{k}\cdot\vgal)t}$, so
that the definition of $\omega$ is independent of $\vgal$, and
corresponds to physical intuition. This
results in the following expressions
\begin{subequations}
\begin{align}
\vec{\mc{E}}^n(\vec{k}) &= \vec{\mc{E}}(\vec{k})
e^{-i(\omega-\vec{k}\cdot\vgal )n \Delta t} \label{eq:app-eigenmodeE}\\
\vec{\mc{B}}^n(\vec{k}) &= \vec{\mc{B}}(\vec{k})
e^{-i(\omega-\vec{k}\cdot\vgal )n \Delta t} \label{eq:app-eigenmodeB}\\
\delta \hat{f}^{n+\frac{1}{2}} \!(\km, \vec{p}) &= \delta \hat{f}(\km,
  \vec{p}) e^{-i(\omega-\vec{k}\cdot\vgal )(n+1/2)\Delta t}
\end{align}
\end{subequations}
Notice that we used $\vec{k}$ instead of $\vec{k}_m \equiv \vec{k} +
\Km$ in the expression of the time evolution of $\delta \hat{f}^{n+\frac{1}{2}} \!(\km,
\vec{p})$. This is because, by definition of an eigenmode, all
quantities (in this case $\vec{\mc{E}}$, $\vec{\mc{B}}$ and $\delta
\hat{f}$) should have the same time evolution.

With these expressions, the discretized Vlasov equation
\cref{eq:app-disc-vlasov} yields:
\begin{equation}
\delta \hat{f}(\km,\vec{p}) = -i\frac{q\Delta t}{2} \mc{S}(\km)\frac{
(\vec{\mc{E}}(\vec{k})
+\vec{v}\times\vec{\mc{B}}(\vec{k}))\cdot\frac{\partial f_0}{\partial
  \vec{p}}}{\sin\left( \frac{(\omega - \vec{k}\cdot\vec{v} -
  \Km\cdot(\vec{v}-\vgal)) \Delta t}{2} \right)}
\end{equation}
And, after some algebra, inserting the above expression into
\cref{eq:app-source-J,eq:app-source-rho1,eq:app-source-rho2} results
in:
\begin{align}
\vec{\mc{J}}^{n+\frac{1}{2}}_d =&\; i\frac{\epsilon_0 \omega_p^2}{\gamma_0}e^{-i(\omega-\vec{k}\cdot\vgal )(n+1/2)\Delta
  t}\nonumber \\
& \times \sum_{\vec{m}} \mc{S}^2(\km) \left(
  \frac{\vec{\mc{F}}}{\frac{2}{\Delta t}s_{\omega'}} +
\frac{c_{\omega'}
  (\km\cdot\vec{\mc{F}})\vec{v}_0}{\left[\frac{2}{\Delta t}
  s_{\omega'}\right]^2}\right) \label{eq:app-expl-J}\\
\theta^*\mc{\rho}^{n+1} - \theta \mc{\rho}^n =&\;
  \frac{2\epsilon_0\omega_p^2}{\gamma_0}s_\omega e^{-i(\omega-\vec{k}\cdot\vgal
  )(n+1/2)\Delta t} \nonumber \\
& \times \sum_{\vec{m}} \mc{S}^2(\km)\frac{
  (\vec{\mc{F}}\cdot\km)}{\left[ \frac{2}{\Delta t}s_{\omega'}\right]^2} \label{eq:app-expl-rho}
\end{align}
where
\begin{align}
&\vec{\mc{F}} \equiv \vec{\mc{E}}(\vec{k}) +
  \vec{v}_0\times\vec{\mc{B}}(\vec{k}) -
  \frac{(\vec{v}_0\cdot\vec{\mc{E}}(\vec{k}))\vec{v}_0}{c^2} \label{eq:app-exprF}\\
& \omega_p^2=\frac{n_0 q^2}{m\epsilon_0}\qquad s_\omega =
  \sin\left(\frac{\omega \Delta t}{2}\right)\\
& s_{\omega'} = \sin\left( \frac{(\omega - \vec{k}\cdot\vec{v}_0 -
  \Km\cdot(\vec{v}_0-\vgal)) \Delta t}{2} \right) \\
& c_{\omega'} = \cos\left( \frac{(\omega - \vec{k}\cdot\vec{v}_0 -
  \Km\cdot(\vec{v}_0-\vgal)) \Delta t}{2} \right)
\end{align}

In the derivation of \cref{eq:app-expl-J,eq:app-expl-rho}, we used
integration by parts and the fact that the distribution function of the unperturbed background plasma, is $f_0(\vec{p}) =
n_0\delta ( \vec{p} - \gamma_0m\vec{v}_0)$, and we also made use of the
relation $\partial_{\vec{p}}\cdot (\vec{\vec{v}}\times
\vec{\mc{B}})=0$ when $\vec{v} = \vec{p}\,(1+(\vec{p}/mc)^2)^{-1/2}/m$.

Finally, inserting
\cref{eq:app-eigenmodeE,eq:app-eigenmodeB,eq:app-expl-J,eq:app-expl-rho}
into the time-symmetrical discrete Maxwell equations
\cref{eq:app-disc-maxwell1,eq:app-disc-maxwell2} results in the
eigensystem:
\begin{subequations}
\begin{align}
s_\omega c\vec{\mc{B}} -t_{ck}
  c_\omega\frac{\vec{k}\times\vec{\mc{E}}}{k} =& -\chi_4' \xi_1
 \frac{\vec{k}\times \vec{\mc{F}}}{k}  
- \chi_4' (\vec{\xi}_2\cdot\vec{\mc{F}}) \frac{\vec{k}\times\vec{v}_0}{ck} \label{eq:app-3deigensyst1} \\
s_\omega \vec{\mc{E}} + t_{ck} c_\omega\frac{\vec{k}\times
  c\vec{\mc{B}}}{k} =& \;s_\omega  (\vec{\xi}_3\cdot\vec{\mc{F}})\frac{\vec{k}}{k}
+ \chi_4 \xi_1 \frac{\vec{k}\times \vec{\mc{F}}}{k}
                       \times\frac{\vec{k}}{k} \nonumber \\
  & + \chi_4 (\vec{\xi}_2\cdot\vec{\mc{F}}) \frac{\vec{k}\times \vec{v}_0}{ck}
                       \times\frac{\vec{k}}{k}  \label{eq:app-3deigensyst2}
\end{align}
\end{subequations}
where the $\xi$ coefficients represent the response of the plasma
\begin{subequations}
\begin{align}
\xi_1 &= \frac{\mc{T} \omega_p^2}{\gamma_0 c k}\left( \sum_{\vec{m}}
\frac{\mc{S}^2(\km)}{\frac{2}{\Delta t}s_{\omega'}} \right) \\
\vec{\xi}_2 &= \frac{\mc{T} \omega_p^2}{\gamma_0 k}\left( \sum_{\vec{m}}
\frac{c_{\omega'}\mc{S}^2(\km)}{\left[\frac{2}{\Delta t}s_{\omega'}\right]^2} \km \right) \\
\vec{\xi}_3 &= \frac{\mc{T} \omega_p^2}{\gamma_0 k}\left( \sum_{\vec{m}}
\frac{\mc{S}^2(\km)}{\left[\frac{2}{\Delta t}s_{\omega'}\right]^2} \km \right) 
\end{align}
\end{subequations}

\subsection{Dispersion relation}

Finally, let us simplify the above eigensystem in the case where both
$\vgal$ and the
velocity of the unperturbed plasma $\vec{v}_0$ are along
$\vec{u}_z$ (i.e. $\vgal = v_{gal}\vec{u}_z$ and $\vec{v}_0 = v_0\vec{u}_z$). In this case, projecting \cref{eq:app-3deigensyst1} along
$y$ and \cref{eq:app-3deigensyst2} along $x$ and $z$ (as well as using the
expression of $\vec{\mc{F}}$ from \cref{eq:app-exprF}) results in the
eigensystem:
\begin{equation} 
M \left( \begin{array}{c}
c \mc{B}_y \\
\mc{E}_z \\
\mc{E}_x 
\end{array}
\right) = \left(
\begin{array}{c}
0 \\
0 \\
0
\end{array}
\right) 
\end{equation}
where the matrix $M$ can be expressed as
\begin{equation} 
M = M_{vacuum} + U_1 (V_1^T + V_2^T) +  s_\omega U_3 V_3^T 
\label{eq:app-exprM}
\end{equation}
where $^T$ denotes the tranpose operation and where
\begin{align}
& M_{vacuum} = \left(
\begin{array}{c c c}
s_\omega & t_{ck}c_\omega k_x/k & -t_{ck}c_\omega k_z/k \\
t_{ck}c_\omega k_x/k & s_\omega & 0 \\
-t_{ck}c_\omega k_z/k & 0 & s_\omega \\
\end{array} \right) 
\\
& U_1 = \left( \begin{array}{c}
\chi_4'\\ \chi_4 k_x/k \\ -\chi_4 k_z/k
\end{array} \right) \quad 
V_1 =\xi_1 \left( \begin{array}{c}
- k_z v_0/(ck) \\ - k_x/(k\gamma_0^2) \\ k_z/k
\end{array} \right) 
\\
& U_2 = \left( \begin{array}{c}
\chi_4\\ \chi_4' k_x/k \\ -\chi_4' k_z/k
\end{array} \right) \quad 
V_2 =\frac{k_x v_0}{ck}\left( \begin{array}{c}
\xi_{2x} v_0/c \\ -\xi_{2z}/\gamma_0^2 \\ -\xi_{2x}
\end{array} \right) \\
&
U_3 = \left( \begin{array}{c}
0 \\ k_z/k \\ k_x/k
\end{array} \right) \quad 
V_3 = \left( \begin{array}{c}
\xi_{3x} v_0/c \\ -\xi_{3z}/\gamma_0^2 \\ -\xi_{3x}
\end{array} \right)  
\end{align}
where we also introduced an additional vector $U_2$ which is not used
in \cref{eq:app-exprM} but will be useful below.

The final dispersion relation is obtained by solving the equation $det(M) =
0$. However, calculating the analytical expression of the determinant
$det(M)$ using e.g. Sarrus' rule can be a daunting
task. Instead, the calculation of $det(M)$ can be faciltated by
expressing the matrix $M$ in the basis $(U_1, U_2, U_3)$ (an operation
which does not change its determinant). In other words, one has
$det(M)=det(M')$ where $M'$ is the expression of the matrix $M$ in the
basis $(U_1, U_2, U_3)$:
\begin{equation}
M' = \left(
\begin{array}{c c c}
s_\omega + (V_1^T+V_2^T)U_1& t_{ck}c_\omega  & s_\omega V_3^TU_1 \\
t_{ck}c_\omega + (V_1^T+V_2^T)U_2 & s_\omega & s_\omega V_3^TU_2 \\
(V_1^T+V_2^T)U_3 & 0 & s_\omega( 1 + V_3^TU_3) \\
\end{array} \right) 
\end{equation}
Using this property, the equation $det(M)=0$ becomes
\begin{align} 
&(  s_\omega^2 - t_{ck}^2 c_\omega^2 ) ( 1 + V_3^T U_3) +
  (V_1^T+V_2^T)U \nonumber \\
& + (V_3^T U_3)( V_1^T U) - (V_1^T U_3)( V_3^T U) \nonumber \\
& + (V_3^T U_3)(V_2^T U) - (V_2^T U_3)( V_3^T U)  = 0
\end{align}
where the trivial solution $s_\omega = 0$ has be discarded, and where
$U = s_\omega U_1 - t_{ck}c_\omega U_2$. After some algebra, this
equation reduces to \cref{eq:dispersion}.

\section{\label{sec:app-cylindricalPSATD}Expression of the PSATD
  equations in quasi-cylindrical geometry}

As mentioned in the text, the PSATD equations in quasi-cylindrical geometry
are obtained from the equations in Cartesian geometry
\cref{eq:continuity,eq:disc-maxwell1,eq:disc-maxwell2}, using the
correspondance table \cref{tab:analogy}. Thus, with this method, the
discretized continuity equation \cref{eq:continuity} becomes
\begin{align}
\label{eq:app-continuity-circ}
-&i k_z v_{gal} \frac{ \mc{\rho}_m^{n+1}  -
 \mc{\rho}_m^n e^{ik_z v_{gal}\Delta t} }{1- e^{ik_z v_{gal}\Delta t}}
   \nonumber \\
&+ k_\perp(\mc{J}^{n+1/2}_{+,m} - \mc{J}^{n+1/2}_{-,m}) + ik_z\mc{J}^{n+1/2}_{z,m} = 0
\end{align}
and the corresponding current correction (\cref{eq:current-correction} in Cartesian
geometry) becomes
\begin{subequations} 
\begin{align}
\mc{J}^{n+1/2}_{+,m} &= \mc{J}^{n+1/2}_{d\;+,m} - k_\perp \mc{G}_m/(2k^2) \\
\mc{J}^{n+1/2}_{-,m} &= \mc{J}^{n+1/2}_{d\;-,m} + k_\perp \mc{G}_m/(2k^2) \\
\mc{J}^{n+1/2}_{z,m} &= \mc{J}^{n+1/2}_{d\;z,m} + i k_z \mc{G}_m/k^2
\end{align}
\end{subequations}
where $k^2= k_\perp^2 + k_z^2$ by definition. In the above equations,
  $\vec{\mc{J}}_d$ and $\vec{\mc{J}}$ are the deposited current and
  corrected current respectively, and the expression of $\mc{G}_m$ is
  obtained by replacing $\vec{\mc{J}}$ by $\vec{\mc{J}}_d$ in the
  left-hand side of \cref{eq:app-continuity-circ}.

Similarly, the update equations for the $\vec{\mc{B}}$ and $\vec{\mc{E}}$
field (\cref{eq:disc-maxwell1,eq:disc-maxwell2}) become:
\begin{widetext}
\begin{subequations}
\begin{align}
\mc{B}^{n+1}_{+,m} &= \theta^2 C \mc{B}_{+,m}^n
 -\frac{\theta^2 S}{ck} \left(k_z\mc{E}^n_{+,m}- \frac{ik_\perp}{2}\mc{E}^n_{z,m} \right)
+ \frac{\theta \chi_1}{\epsilon_0c^2k^2}\left( k_z \mc{J}^{n+1/2}_{+,m} -
\frac{ik_\perp}{2} \mc{J}^{n+1/2}_{z,m} \right) \\
\mc{B}^{n+1}_{-,m} &= \theta^2 C \mc{B}_{-,m}^n
 -\frac{\theta^2 S}{ck}\left( - k_z\mc{E}^n_{-,m}- \frac{ik_\perp}{2}\mc{E}^n_{z,m} \right)
+ \;\frac{\theta \chi_1}{\epsilon_0c^2k^2}\left( - k_z \mc{J}^{n+1/2}_{-,m} -
\frac{ik_\perp}{2} \mc{J}^{n+1/2}_{z,m} \right)\\
\mc{B}^{n+1}_{z,m} &= \theta^2 C \mc{B}^n_{z,m}
 -\frac{\theta^2 S}{ck}\left( ik_\perp \mc{E}^{n}_{+,m} +
ik_\perp \mc{E}^{n}_{-,m} \right)
+ \;\frac{\theta \chi_1}{\epsilon_0c^2k^2}\left( ik_\perp \mc{J}^{n+1/2}_{+,m} +
ik_\perp \mc{J}^{n+1/2}_{-,m} \right)
\end{align}
\end{subequations}
\begin{subequations}
\begin{align}
\mc{E}^{n+1}_{+,m} &=  \theta^2 C  \mc{E}^n_{+,m}
 +\frac{\theta^2 S c}{k} \left(k_z\mc{B}^n_{+,m}- \frac{ik_\perp}{2}\mc{B}^n_{z,m} \right)
+\frac{i\nu \theta \chi_1 - \theta^2S}{\epsilon_0 ck} \; \mc{J}_{+,m}^{n+1/2}
 + \frac{1}{\epsilon_0k^2}\left(\; \chi_2\;\mc{\rho}_m^{n+1} -
  \theta^2\chi_3\;\mc{\rho}_m^{n} \;\right) \frac{k_\perp}{2} \\
\mc{E}^{n+1}_{-,m} &=  \theta^2 C  \mc{E}^n_{-,m}
 +\frac{\theta^2 S c}{k} \left( - k_z\mc{B}^n_{-,m}- \frac{ik_\perp}{2}\mc{B}^n_{z,m} \right)
+\frac{i\nu \theta \chi_1 - \theta^2S}{\epsilon_0 ck} \; \mc{J}_{-,m}^{n+1/2}
 - \frac{1}{\epsilon_0k^2}\left(\; \chi_2\;\mc{\rho}_m^{n+1} -
  \theta^2\chi_3\;\mc{\rho}_m^{n} \;\right) \frac{k_\perp}{2} \\
\mc{E}^{n+1}_{z,m} &=  \theta^2 C  \mc{E}^n_{z,m}
 +\frac{\theta^2 S c}{k} \left( ik_\perp \mc{B}^{n}_{+,m} +
ik_\perp \mc{B}^{n}_{-,m} \right) 
+\frac{i\nu \theta \chi_1 - \theta^2S}{\epsilon_0 ck} \; \mc{J}_{z,m}^{n+1/2}
 - \frac{1}{\epsilon_0k^2}\left(\; \chi_2\;\mc{\rho}_m^{n+1} -
  \theta^2\chi_3\;\mc{\rho}_m^{n} \;\right) i k_z
\end{align}
\end{subequations}
\end{widetext}
In the above equations, the coefficients $\theta$,
$C$, $S$, $\nu$, $\chi_1$, $\chi_2$ and $\chi_3$ have the same
expression as in
\cref{eq:def-C-S,eq:def-nu-theta,eq:def-chi1,eq:def-chi23} 
(bearing in my mind, in quasi-cylindrical geometry, that $\vgal$ is
necessarily along $\vec{u}_z$ and
that the expression of $k$ is $\sqrt{k_\perp^2 + k_z^2}$).

\bibliography{Bibliography.bib}

\begin{thebibliography}{36}%
\makeatletter
\providecommand \@ifxundefined [1]{%
 \@ifx{#1\undefined}
}%
\providecommand \@ifnum [1]{%
 \ifnum #1\expandafter \@firstoftwo
 \else \expandafter \@secondoftwo
 \fi
}%
\providecommand \@ifx [1]{%
 \ifx #1\expandafter \@firstoftwo
 \else \expandafter \@secondoftwo
 \fi
}%
\providecommand \natexlab [1]{#1}%
\providecommand \enquote  [1]{``#1''}%
\providecommand \bibnamefont  [1]{#1}%
\providecommand \bibfnamefont [1]{#1}%
\providecommand \citenamefont [1]{#1}%
\providecommand \href@noop [0]{\@secondoftwo}%
\providecommand \href [0]{\begingroup \@sanitize@url \@href}%
\providecommand \@href[1]{\@@startlink{#1}\@@href}%
\providecommand \@@href[1]{\endgroup#1\@@endlink}%
\providecommand \@sanitize@url [0]{\catcode `\\12\catcode `\$12\catcode
  `\&12\catcode `\#12\catcode `\^12\catcode `\_12\catcode `\%12\relax}%
\providecommand \@@startlink[1]{}%
\providecommand \@@endlink[0]{}%
\providecommand \url  [0]{\begingroup\@sanitize@url \@url }%
\providecommand \@url [1]{\endgroup\@href {#1}{\urlprefix }}%
\providecommand \urlprefix  [0]{URL }%
\providecommand \Eprint [0]{\href }%
\providecommand \doibase [0]{http://dx.doi.org/}%
\providecommand \selectlanguage [0]{\@gobble}%
\providecommand \bibinfo  [0]{\@secondoftwo}%
\providecommand \bibfield  [0]{\@secondoftwo}%
\providecommand \translation [1]{[#1]}%
\providecommand \BibitemOpen [0]{}%
\providecommand \bibitemStop [0]{}%
\providecommand \bibitemNoStop [0]{.\EOS\space}%
\providecommand \EOS [0]{\spacefactor3000\relax}%
\providecommand \BibitemShut  [1]{\csname bibitem#1\endcsname}%
\let\auto@bib@innerbib\@empty
\bibitem [{\citenamefont {{Spitkovsky}}(2008)}]{SpitkovskyApJ2008}%
  \BibitemOpen
  \bibfield  {author} {\bibinfo {author} {\bibfnamefont {A.}~\bibnamefont
  {{Spitkovsky}}},\ }\href {\doibase 10.1086/590248} {\bibfield  {journal}
  {\bibinfo  {journal} {The Astrophysical Journal Letters}\ }\textbf {\bibinfo
  {volume} {682}},\ \bibinfo {pages} {L5} (\bibinfo {year} {2008})},\ \Eprint
  {http://arxiv.org/abs/0802.3216} {0802.3216} \BibitemShut {NoStop}%
\bibitem [{\citenamefont {Keshet}\ \emph {et~al.}(2009)\citenamefont {Keshet},
  \citenamefont {Katz}, \citenamefont {Spitkovsky},\ and\ \citenamefont
  {Waxman}}]{KeshetApJ2009}%
  \BibitemOpen
  \bibfield  {author} {\bibinfo {author} {\bibfnamefont {U.}~\bibnamefont
  {Keshet}}, \bibinfo {author} {\bibfnamefont {B.}~\bibnamefont {Katz}},
  \bibinfo {author} {\bibfnamefont {A.}~\bibnamefont {Spitkovsky}}, \ and\
  \bibinfo {author} {\bibfnamefont {E.}~\bibnamefont {Waxman}},\ }\href
  {http://stacks.iop.org/1538-4357/693/i=2/a=L127} {\bibfield  {journal}
  {\bibinfo  {journal} {The Astrophysical Journal Letters}\ }\textbf {\bibinfo
  {volume} {693}},\ \bibinfo {pages} {L127} (\bibinfo {year}
  {2009})}\BibitemShut {NoStop}%
\bibitem [{\citenamefont {Tajima}\ and\ \citenamefont
  {Dawson}(1979)}]{TajimaPRL1979}%
  \BibitemOpen
  \bibfield  {author} {\bibinfo {author} {\bibfnamefont {T.}~\bibnamefont
  {Tajima}}\ and\ \bibinfo {author} {\bibfnamefont {J.~M.}\ \bibnamefont
  {Dawson}},\ }\href {\doibase 10.1103/PhysRevLett.43.267} {\bibfield
  {journal} {\bibinfo  {journal} {Phys. Rev. Lett.}\ }\textbf {\bibinfo
  {volume} {43}},\ \bibinfo {pages} {267} (\bibinfo {year} {1979})}\BibitemShut
  {NoStop}%
\bibitem [{\citenamefont {Vay}(2007)}]{VayPRL2007}%
  \BibitemOpen
  \bibfield  {author} {\bibinfo {author} {\bibfnamefont {J.-L.}\ \bibnamefont
  {Vay}},\ }\href {\doibase 10.1103/PhysRevLett.98.130405} {\bibfield
  {journal} {\bibinfo  {journal} {Phys. Rev. Lett.}\ }\textbf {\bibinfo
  {volume} {98}},\ \bibinfo {pages} {130405} (\bibinfo {year}
  {2007})}\BibitemShut {NoStop}%
\bibitem [{\citenamefont {Hockney}\ and\ \citenamefont
  {Eastwood}(1988)}]{Hockney1988}%
  \BibitemOpen
  \bibfield  {author} {\bibinfo {author} {\bibfnamefont {R.}~\bibnamefont
  {Hockney}}\ and\ \bibinfo {author} {\bibfnamefont {J.}~\bibnamefont
  {Eastwood}},\ }\href {http://books.google.fr/books?id=nTOFkmnCQuIC} {\emph
  {\bibinfo {title} {Computer Simulation Using Particles}}}\ (\bibinfo
  {publisher} {Taylor \& Francis},\ \bibinfo {year} {1988})\BibitemShut
  {NoStop}%
\bibitem [{\citenamefont {Birdsall}\ and\ \citenamefont
  {Langdon}(2004{\natexlab{a}})}]{Birdsall2004}%
  \BibitemOpen
  \bibfield  {author} {\bibinfo {author} {\bibfnamefont {C.}~\bibnamefont
  {Birdsall}}\ and\ \bibinfo {author} {\bibfnamefont {A.}~\bibnamefont
  {Langdon}},\ }\href {http://books.google.fr/books?id=S2lqgDTm6a4C} {\emph
  {\bibinfo {title} {Plasma Physics via Computer Simulation, Appendix E}}},\
  Series in Plasma Physics\ (\bibinfo  {publisher} {Taylor \& Francis},\
  \bibinfo {year} {2004})\BibitemShut {NoStop}%
\bibitem [{\citenamefont {Martins}\ \emph {et~al.}(2010)\citenamefont
  {Martins}, \citenamefont {Fonseca}, \citenamefont {Silva}, \citenamefont
  {Lu},\ and\ \citenamefont {Mori}}]{MartinsCPC10}%
  \BibitemOpen
  \bibfield  {author} {\bibinfo {author} {\bibfnamefont {S.~F.}\ \bibnamefont
  {Martins}}, \bibinfo {author} {\bibfnamefont {R.~A.}\ \bibnamefont
  {Fonseca}}, \bibinfo {author} {\bibfnamefont {L.~O.}\ \bibnamefont {Silva}},
  \bibinfo {author} {\bibfnamefont {W.}~\bibnamefont {Lu}}, \ and\ \bibinfo
  {author} {\bibfnamefont {W.~B.}\ \bibnamefont {Mori}},\ }\href {\doibase
  http://dx.doi.org/10.1016/j.cpc.2009.12.023} {\bibfield  {journal} {\bibinfo
  {journal} {Computer Physics Communications}\ }\textbf {\bibinfo {volume}
  {181}},\ \bibinfo {pages} {869 } (\bibinfo {year} {2010})}\BibitemShut
  {NoStop}%
\bibitem [{\citenamefont {Vay}\ \emph {et~al.}(2010)\citenamefont {Vay},
  \citenamefont {Geddes}, \citenamefont {Benedetti}, \citenamefont {Bruhwiler},
  \citenamefont {Cormier‐Michel}, \citenamefont {Cowan}, \citenamefont
  {Cary},\ and\ \citenamefont {Grote}}]{VayAAC10}%
  \BibitemOpen
  \bibfield  {author} {\bibinfo {author} {\bibfnamefont {J.}~\bibnamefont
  {Vay}}, \bibinfo {author} {\bibfnamefont {C.~G.~R.}\ \bibnamefont {Geddes}},
  \bibinfo {author} {\bibfnamefont {C.}~\bibnamefont {Benedetti}}, \bibinfo
  {author} {\bibfnamefont {D.~L.}\ \bibnamefont {Bruhwiler}}, \bibinfo {author}
  {\bibfnamefont {E.}~\bibnamefont {Cormier‐Michel}}, \bibinfo {author}
  {\bibfnamefont {B.~M.}\ \bibnamefont {Cowan}}, \bibinfo {author}
  {\bibfnamefont {J.~R.}\ \bibnamefont {Cary}}, \ and\ \bibinfo {author}
  {\bibfnamefont {D.~P.}\ \bibnamefont {Grote}},\ }\href {\doibase
  http://dx.doi.org/10.1063/1.3520322} {\bibfield  {journal} {\bibinfo
  {journal} {AIP Conference Proceedings}\ }\textbf {\bibinfo {volume} {1299}},\
  \bibinfo {pages} {244} (\bibinfo {year} {2010})}\BibitemShut {NoStop}%
\bibitem [{\citenamefont {Vay}\ \emph {et~al.}(2011{\natexlab{a}})\citenamefont
  {Vay}, \citenamefont {Geddes}, \citenamefont {Cormier-Michel},\ and\
  \citenamefont {Grote}}]{VayJCP11}%
  \BibitemOpen
  \bibfield  {author} {\bibinfo {author} {\bibfnamefont {J.-L.}\ \bibnamefont
  {Vay}}, \bibinfo {author} {\bibfnamefont {C.}~\bibnamefont {Geddes}},
  \bibinfo {author} {\bibfnamefont {E.}~\bibnamefont {Cormier-Michel}}, \ and\
  \bibinfo {author} {\bibfnamefont {D.}~\bibnamefont {Grote}},\ }\href
  {\doibase http://dx.doi.org/10.1016/j.jcp.2011.04.003} {\bibfield  {journal}
  {\bibinfo  {journal} {Journal of Computational Physics}\ }\textbf {\bibinfo
  {volume} {230}},\ \bibinfo {pages} {5908 } (\bibinfo {year}
  {2011}{\natexlab{a}})}\BibitemShut {NoStop}%
\bibitem [{\citenamefont {Vay}\ \emph {et~al.}(2011{\natexlab{b}})\citenamefont
  {Vay}, \citenamefont {Geddes}, \citenamefont {Cormier-Michel},\ and\
  \citenamefont {Grote}}]{VayPOPL11}%
  \BibitemOpen
  \bibfield  {author} {\bibinfo {author} {\bibfnamefont {J.-L.}\ \bibnamefont
  {Vay}}, \bibinfo {author} {\bibfnamefont {C.~G.~R.}\ \bibnamefont {Geddes}},
  \bibinfo {author} {\bibfnamefont {E.}~\bibnamefont {Cormier-Michel}}, \ and\
  \bibinfo {author} {\bibfnamefont {D.~P.}\ \bibnamefont {Grote}},\ }\href
  {\doibase http://dx.doi.org/10.1063/1.3559483} {\bibfield  {journal}
  {\bibinfo  {journal} {Physics of Plasmas}\ }\textbf {\bibinfo {volume}
  {18}},\ \bibinfo {eid} {030701} (\bibinfo {year} {2011}{\natexlab{b}}),\
  http://dx.doi.org/10.1063/1.3559483}\BibitemShut {NoStop}%
\bibitem [{\citenamefont {Godfrey}(1974)}]{GodfreyJCP1974}%
  \BibitemOpen
  \bibfield  {author} {\bibinfo {author} {\bibfnamefont {B.~B.}\ \bibnamefont
  {Godfrey}},\ }\href {\doibase 10.1016/0021-9991(74)90076-X} {\bibfield
  {journal} {\bibinfo  {journal} {Journal of Computational Physics}\ }\textbf
  {\bibinfo {volume} {15}},\ \bibinfo {pages} {504 } (\bibinfo {year}
  {1974})}\BibitemShut {NoStop}%
\bibitem [{\citenamefont {Godfrey}(1975)}]{GodfreyJCP1975}%
  \BibitemOpen
  \bibfield  {author} {\bibinfo {author} {\bibfnamefont {B.~B.}\ \bibnamefont
  {Godfrey}},\ }\href {\doibase 10.1016/0021-9991(75)90116-3} {\bibfield
  {journal} {\bibinfo  {journal} {Journal of Computational Physics}\ }\textbf
  {\bibinfo {volume} {19}},\ \bibinfo {pages} {58 } (\bibinfo {year}
  {1975})}\BibitemShut {NoStop}%
\bibitem [{\citenamefont {Godfrey}\ and\ \citenamefont
  {Vay}(2013)}]{GodfreyJCP2013}%
  \BibitemOpen
  \bibfield  {author} {\bibinfo {author} {\bibfnamefont {B.~B.}\ \bibnamefont
  {Godfrey}}\ and\ \bibinfo {author} {\bibfnamefont {J.-L.}\ \bibnamefont
  {Vay}},\ }\href {\doibase 10.1016/j.jcp.2013.04.006} {\bibfield  {journal}
  {\bibinfo  {journal} {Journal of Computational Physics}\ }\textbf {\bibinfo
  {volume} {248}},\ \bibinfo {pages} {33 } (\bibinfo {year}
  {2013})}\BibitemShut {NoStop}%
\bibitem [{\citenamefont {Xu}\ \emph {et~al.}(2013)\citenamefont {Xu},
  \citenamefont {Yu}, \citenamefont {Martins}, \citenamefont {Tsung},
  \citenamefont {Decyk}, \citenamefont {Vieira}, \citenamefont {Fonseca},
  \citenamefont {Lu}, \citenamefont {Silva},\ and\ \citenamefont
  {Mori}}]{XuCPC2013}%
  \BibitemOpen
  \bibfield  {author} {\bibinfo {author} {\bibfnamefont {X.}~\bibnamefont
  {Xu}}, \bibinfo {author} {\bibfnamefont {P.}~\bibnamefont {Yu}}, \bibinfo
  {author} {\bibfnamefont {S.~F.}\ \bibnamefont {Martins}}, \bibinfo {author}
  {\bibfnamefont {F.~S.}\ \bibnamefont {Tsung}}, \bibinfo {author}
  {\bibfnamefont {V.~K.}\ \bibnamefont {Decyk}}, \bibinfo {author}
  {\bibfnamefont {J.}~\bibnamefont {Vieira}}, \bibinfo {author} {\bibfnamefont
  {R.~A.}\ \bibnamefont {Fonseca}}, \bibinfo {author} {\bibfnamefont
  {W.}~\bibnamefont {Lu}}, \bibinfo {author} {\bibfnamefont {L.~O.}\
  \bibnamefont {Silva}}, \ and\ \bibinfo {author} {\bibfnamefont {W.~B.}\
  \bibnamefont {Mori}},\ }\href {\doibase 10.1016/j.cpc.2013.07.003} {\bibfield
   {journal} {\bibinfo  {journal} {Computer Physics Communications}\ }\textbf
  {\bibinfo {volume} {184}},\ \bibinfo {pages} {2503 } (\bibinfo {year}
  {2013})}\BibitemShut {NoStop}%
\bibitem [{\citenamefont {Godfrey}\ \emph
  {et~al.}(2014{\natexlab{a}})\citenamefont {Godfrey}, \citenamefont {Vay},\
  and\ \citenamefont {Haber}}]{GodfreyJCP2014}%
  \BibitemOpen
  \bibfield  {author} {\bibinfo {author} {\bibfnamefont {B.~B.}\ \bibnamefont
  {Godfrey}}, \bibinfo {author} {\bibfnamefont {J.-L.}\ \bibnamefont {Vay}}, \
  and\ \bibinfo {author} {\bibfnamefont {I.}~\bibnamefont {Haber}},\ }\href
  {\doibase 10.1016/j.jcp.2013.10.053} {\bibfield  {journal} {\bibinfo
  {journal} {Journal of Computational Physics}\ }\textbf {\bibinfo {volume}
  {258}},\ \bibinfo {pages} {689 } (\bibinfo {year}
  {2014}{\natexlab{a}})}\BibitemShut {NoStop}%
\bibitem [{\citenamefont {Godfrey}\ \emph
  {et~al.}(2014{\natexlab{b}})\citenamefont {Godfrey}, \citenamefont {Vay},\
  and\ \citenamefont {Haber}}]{GodfreyIEEE2014}%
  \BibitemOpen
  \bibfield  {author} {\bibinfo {author} {\bibfnamefont {B.}~\bibnamefont
  {Godfrey}}, \bibinfo {author} {\bibfnamefont {J.-L.}\ \bibnamefont {Vay}}, \
  and\ \bibinfo {author} {\bibfnamefont {I.}~\bibnamefont {Haber}},\ }\href
  {\doibase 10.1109/TPS.2014.2310654} {\bibfield  {journal} {\bibinfo
  {journal} {Plasma Science, IEEE Transactions on}\ }\textbf {\bibinfo {volume}
  {42}},\ \bibinfo {pages} {1339} (\bibinfo {year}
  {2014}{\natexlab{b}})}\BibitemShut {NoStop}%
\bibitem [{\citenamefont {Godfrey}\ and\ \citenamefont
  {Vay}(2014)}]{GodfreyJCP2014b}%
  \BibitemOpen
  \bibfield  {author} {\bibinfo {author} {\bibfnamefont {B.~B.}\ \bibnamefont
  {Godfrey}}\ and\ \bibinfo {author} {\bibfnamefont {J.-L.}\ \bibnamefont
  {Vay}},\ }\href {\doibase 10.1016/j.jcp.2014.02.022} {\bibfield  {journal}
  {\bibinfo  {journal} {Journal of Computational Physics}\ }\textbf {\bibinfo
  {volume} {267}},\ \bibinfo {pages} {1 } (\bibinfo {year} {2014})}\BibitemShut
  {NoStop}%
\bibitem [{\citenamefont {{Godfrey}}(2014)}]{Godfreyarxiv2014}%
  \BibitemOpen
  \bibfield  {author} {\bibinfo {author} {\bibfnamefont {B.~B.}\ \bibnamefont
  {{Godfrey}}},\ }\href@noop {} {\bibfield  {journal} {\bibinfo  {journal}
  {ArXiv e-prints}\ } (\bibinfo {year} {2014})},\ \Eprint
  {http://arxiv.org/abs/1408.1146} {arXiv:1408.1146 [physics.plasm-ph]}
  \BibitemShut {NoStop}%
\bibitem [{\citenamefont {Godfrey}\ and\ \citenamefont
  {Vay}(2015)}]{GodfreyCPC2015}%
  \BibitemOpen
  \bibfield  {author} {\bibinfo {author} {\bibfnamefont {B.~B.}\ \bibnamefont
  {Godfrey}}\ and\ \bibinfo {author} {\bibfnamefont {J.-L.}\ \bibnamefont
  {Vay}},\ }\href {\doibase 10.1016/j.cpc.2015.06.008} {\bibfield  {journal}
  {\bibinfo  {journal} {Computer Physics Communications}\ ,\ } (\bibinfo {year}
  {2015})}\BibitemShut {NoStop}%
\bibitem [{\citenamefont {Yu}\ \emph {et~al.}(2015{\natexlab{a}})\citenamefont
  {Yu}, \citenamefont {Xu}, \citenamefont {Decyk}, \citenamefont {Fiuza},
  \citenamefont {Vieira}, \citenamefont {Tsung}, \citenamefont {Fonseca},
  \citenamefont {Lu}, \citenamefont {Silva},\ and\ \citenamefont
  {Mori}}]{YuCPC2015}%
  \BibitemOpen
  \bibfield  {author} {\bibinfo {author} {\bibfnamefont {P.}~\bibnamefont
  {Yu}}, \bibinfo {author} {\bibfnamefont {X.}~\bibnamefont {Xu}}, \bibinfo
  {author} {\bibfnamefont {V.~K.}\ \bibnamefont {Decyk}}, \bibinfo {author}
  {\bibfnamefont {F.}~\bibnamefont {Fiuza}}, \bibinfo {author} {\bibfnamefont
  {J.}~\bibnamefont {Vieira}}, \bibinfo {author} {\bibfnamefont {F.~S.}\
  \bibnamefont {Tsung}}, \bibinfo {author} {\bibfnamefont {R.~A.}\ \bibnamefont
  {Fonseca}}, \bibinfo {author} {\bibfnamefont {W.}~\bibnamefont {Lu}},
  \bibinfo {author} {\bibfnamefont {L.~O.}\ \bibnamefont {Silva}}, \ and\
  \bibinfo {author} {\bibfnamefont {W.~B.}\ \bibnamefont {Mori}},\ }\href
  {\doibase 10.1016/j.cpc.2015.02.018} {\bibfield  {journal} {\bibinfo
  {journal} {Computer Physics Communications}\ }\textbf {\bibinfo {volume}
  {192}},\ \bibinfo {pages} {32 } (\bibinfo {year}
  {2015}{\natexlab{a}})}\BibitemShut {NoStop}%
\bibitem [{\citenamefont {Yu}\ \emph {et~al.}(2015{\natexlab{b}})\citenamefont
  {Yu}, \citenamefont {Xu}, \citenamefont {Tableman}, \citenamefont {Decyk},
  \citenamefont {Tsung}, \citenamefont {Fiuza}, \citenamefont {Davidson},
  \citenamefont {Vieira}, \citenamefont {Fonseca}, \citenamefont {Lu},
  \citenamefont {Silva},\ and\ \citenamefont {Mori}}]{YuCPC2015-Circ}%
  \BibitemOpen
  \bibfield  {author} {\bibinfo {author} {\bibfnamefont {P.}~\bibnamefont
  {Yu}}, \bibinfo {author} {\bibfnamefont {X.}~\bibnamefont {Xu}}, \bibinfo
  {author} {\bibfnamefont {A.}~\bibnamefont {Tableman}}, \bibinfo {author}
  {\bibfnamefont {V.~K.}\ \bibnamefont {Decyk}}, \bibinfo {author}
  {\bibfnamefont {F.~S.}\ \bibnamefont {Tsung}}, \bibinfo {author}
  {\bibfnamefont {F.}~\bibnamefont {Fiuza}}, \bibinfo {author} {\bibfnamefont
  {A.}~\bibnamefont {Davidson}}, \bibinfo {author} {\bibfnamefont
  {J.}~\bibnamefont {Vieira}}, \bibinfo {author} {\bibfnamefont {R.~A.}\
  \bibnamefont {Fonseca}}, \bibinfo {author} {\bibfnamefont {W.}~\bibnamefont
  {Lu}}, \bibinfo {author} {\bibfnamefont {L.~O.}\ \bibnamefont {Silva}}, \
  and\ \bibinfo {author} {\bibfnamefont {W.~B.}\ \bibnamefont {Mori}},\ }\href
  {\doibase 10.1016/j.cpc.2015.08.026} {\bibfield  {journal} {\bibinfo
  {journal} {Computer Physics Communications}\ }\textbf {\bibinfo {volume}
  {197}},\ \bibinfo {pages} {144 } (\bibinfo {year}
  {2015}{\natexlab{b}})}\BibitemShut {NoStop}%
\bibitem [{\citenamefont {{Li}}\ \emph {et~al.}(2016)\citenamefont {{Li}},
  \citenamefont {{Yu}}, \citenamefont {{Xu}}, \citenamefont {{Fiuza}},
  \citenamefont {{Decyk}}, \citenamefont {{Dalichaouch}}, \citenamefont
  {{Davidson}}, \citenamefont {{Tableman}}, \citenamefont {{An}}, \citenamefont
  {{Tsung}}, \citenamefont {{Fonseca}}, \citenamefont {{Lu}},\ and\
  \citenamefont {{Mori}}}]{Yu-arxiv2016}%
  \BibitemOpen
  \bibfield  {author} {\bibinfo {author} {\bibfnamefont {F.}~\bibnamefont
  {{Li}}}, \bibinfo {author} {\bibfnamefont {P.}~\bibnamefont {{Yu}}}, \bibinfo
  {author} {\bibfnamefont {X.}~\bibnamefont {{Xu}}}, \bibinfo {author}
  {\bibfnamefont {F.}~\bibnamefont {{Fiuza}}}, \bibinfo {author} {\bibfnamefont
  {V.~K.}\ \bibnamefont {{Decyk}}}, \bibinfo {author} {\bibfnamefont
  {T.}~\bibnamefont {{Dalichaouch}}}, \bibinfo {author} {\bibfnamefont
  {A.}~\bibnamefont {{Davidson}}}, \bibinfo {author} {\bibfnamefont
  {A.}~\bibnamefont {{Tableman}}}, \bibinfo {author} {\bibfnamefont
  {W.}~\bibnamefont {{An}}}, \bibinfo {author} {\bibfnamefont {F.~S.}\
  \bibnamefont {{Tsung}}}, \bibinfo {author} {\bibfnamefont {R.~A.}\
  \bibnamefont {{Fonseca}}}, \bibinfo {author} {\bibfnamefont {W.}~\bibnamefont
  {{Lu}}}, \ and\ \bibinfo {author} {\bibfnamefont {W.~B.}\ \bibnamefont
  {{Mori}}},\ }\href@noop {} {\bibfield  {journal} {\bibinfo  {journal} {ArXiv
  e-prints}\ } (\bibinfo {year} {2016})},\ \Eprint
  {http://arxiv.org/abs/1605.01496} {arXiv:1605.01496 [physics.comp-ph]}
  \BibitemShut {NoStop}%
\bibitem [{\citenamefont {Lifschitz}\ \emph {et~al.}(2009)\citenamefont
  {Lifschitz}, \citenamefont {Davoine}, \citenamefont {Lefebvre}, \citenamefont
  {Faure}, \citenamefont {Rechatin},\ and\ \citenamefont {Malka}}]{Lifschitz}%
  \BibitemOpen
  \bibfield  {author} {\bibinfo {author} {\bibfnamefont {A.~F.}\ \bibnamefont
  {Lifschitz}}, \bibinfo {author} {\bibfnamefont {X.}~\bibnamefont {Davoine}},
  \bibinfo {author} {\bibfnamefont {E.}~\bibnamefont {Lefebvre}}, \bibinfo
  {author} {\bibfnamefont {J.}~\bibnamefont {Faure}}, \bibinfo {author}
  {\bibfnamefont {C.}~\bibnamefont {Rechatin}}, \ and\ \bibinfo {author}
  {\bibfnamefont {V.}~\bibnamefont {Malka}},\ }\href {\doibase
  10.1016/j.jcp.2008.11.017} {\bibfield  {journal} {\bibinfo  {journal} {J.
  Comput. Phys.}\ }\textbf {\bibinfo {volume} {228}},\ \bibinfo {pages} {1803}
  (\bibinfo {year} {2009})}\BibitemShut {NoStop}%
\bibitem [{\citenamefont {{Davidson}}\ \emph {et~al.}(2015)\citenamefont
  {{Davidson}}, \citenamefont {{Tableman}}, \citenamefont {{An}}, \citenamefont
  {{Tsung}}, \citenamefont {{Lu}}, \citenamefont {{Vieira}}, \citenamefont
  {{Fonseca}}, \citenamefont {{Silva}},\ and\ \citenamefont
  {{Mori}}}]{Davidson}%
  \BibitemOpen
  \bibfield  {author} {\bibinfo {author} {\bibfnamefont {A.}~\bibnamefont
  {{Davidson}}}, \bibinfo {author} {\bibfnamefont {A.}~\bibnamefont
  {{Tableman}}}, \bibinfo {author} {\bibfnamefont {W.}~\bibnamefont {{An}}},
  \bibinfo {author} {\bibfnamefont {F.~S.}\ \bibnamefont {{Tsung}}}, \bibinfo
  {author} {\bibfnamefont {W.}~\bibnamefont {{Lu}}}, \bibinfo {author}
  {\bibfnamefont {J.}~\bibnamefont {{Vieira}}}, \bibinfo {author}
  {\bibfnamefont {R.~A.}\ \bibnamefont {{Fonseca}}}, \bibinfo {author}
  {\bibfnamefont {L.~O.}\ \bibnamefont {{Silva}}}, \ and\ \bibinfo {author}
  {\bibfnamefont {W.~B.}\ \bibnamefont {{Mori}}},\ }\href {\doibase
  10.1016/j.jcp.2014.10.064} {\bibfield  {journal} {\bibinfo  {journal}
  {Journal of Computational Physics}\ }\textbf {\bibinfo {volume} {281}},\
  \bibinfo {pages} {1063} (\bibinfo {year} {2015})},\ \Eprint
  {http://arxiv.org/abs/1403.6890} {arXiv:1403.6890 [physics.comp-ph]}
  \BibitemShut {NoStop}%
\bibitem [{\citenamefont {Kirchen}\ \emph {et~al.}(2016)\citenamefont
  {Kirchen}, \citenamefont {Lehe}, \citenamefont {Godfrey}, \citenamefont
  {Vay},\ and\ \citenamefont {Maier}}]{Kirchen2016}%
  \BibitemOpen
  \bibfield  {author} {\bibinfo {author} {\bibfnamefont {M.}~\bibnamefont
  {Kirchen}}, \bibinfo {author} {\bibfnamefont {R.}~\bibnamefont {Lehe}},
  \bibinfo {author} {\bibfnamefont {B.~B.}\ \bibnamefont {Godfrey}}, \bibinfo
  {author} {\bibfnamefont {J.-L.}\ \bibnamefont {Vay}}, \ and\ \bibinfo
  {author} {\bibfnamefont {A.~R.}\ \bibnamefont {Maier}},\ }\href@noop {}
  {\bibfield  {journal} {\bibinfo  {journal} {to be submitted}\ } (\bibinfo
  {year} {2016})}\BibitemShut {NoStop}%
\bibitem [{\citenamefont {Haber}\ \emph {et~al.}(1973)\citenamefont {Haber},
  \citenamefont {Lee}, \citenamefont {Klein},\ and\ \citenamefont
  {Boris}}]{Haber}%
  \BibitemOpen
  \bibfield  {author} {\bibinfo {author} {\bibfnamefont {I.}~\bibnamefont
  {Haber}}, \bibinfo {author} {\bibfnamefont {R.}~\bibnamefont {Lee}}, \bibinfo
  {author} {\bibfnamefont {H.}~\bibnamefont {Klein}}, \ and\ \bibinfo {author}
  {\bibfnamefont {J.}~\bibnamefont {Boris}},\ }\href@noop {} {\emph {\bibinfo
  {title} {Proc. Sixth Conf. on Num. Sim. Plasmas, Berkeley, CA}}}\ (\bibinfo
  {year} {1973})\BibitemShut {NoStop}%
\bibitem [{\citenamefont {Vay}\ \emph {et~al.}(2013)\citenamefont {Vay},
  \citenamefont {Haber},\ and\ \citenamefont {Godfrey}}]{VayJCP2013}%
  \BibitemOpen
  \bibfield  {author} {\bibinfo {author} {\bibfnamefont {J.-L.}\ \bibnamefont
  {Vay}}, \bibinfo {author} {\bibfnamefont {I.}~\bibnamefont {Haber}}, \ and\
  \bibinfo {author} {\bibfnamefont {B.~B.}\ \bibnamefont {Godfrey}},\ }\href
  {\doibase 10.1016/j.jcp.2013.03.010} {\bibfield  {journal} {\bibinfo
  {journal} {Journal of Computational Physics}\ }\textbf {\bibinfo {volume}
  {243}},\ \bibinfo {pages} {260 } (\bibinfo {year} {2013})}\BibitemShut
  {NoStop}%
\bibitem [{\citenamefont {Boris}(1970)}]{Boris1970}%
  \BibitemOpen
  \bibfield  {author} {\bibinfo {author} {\bibfnamefont {J.}~\bibnamefont
  {Boris}},\ }in\ \href@noop {} {\emph {\bibinfo {booktitle} {Proceeding of the
  Fourth Conference on Numerical Simulations of Plasmas}}}\ (\bibinfo
  {publisher} {Naval Research Laboratory},\ \bibinfo {year} {1970})\BibitemShut
  {NoStop}%
\bibitem [{\citenamefont {Vay}(2008)}]{VayPoP2008}%
  \BibitemOpen
  \bibfield  {author} {\bibinfo {author} {\bibfnamefont {J.-L.}\ \bibnamefont
  {Vay}},\ }\href {\doibase 10.1063/1.2837054} {\bibfield  {journal} {\bibinfo
  {journal} {Physics of Plasmas (1994-present)}\ }\textbf {\bibinfo {volume}
  {15}},\ \bibinfo {eid} {056701} (\bibinfo {year} {2008})}\BibitemShut
  {NoStop}%
\bibitem [{\citenamefont {Vay}\ \emph {et~al.}(2012)\citenamefont {Vay},
  \citenamefont {Grote}, \citenamefont {Cohen},\ and\ \citenamefont
  {Friedman}}]{Warpref}%
  \BibitemOpen
  \bibfield  {author} {\bibinfo {author} {\bibfnamefont {J.-L.}\ \bibnamefont
  {Vay}}, \bibinfo {author} {\bibfnamefont {D.~P.}\ \bibnamefont {Grote}},
  \bibinfo {author} {\bibfnamefont {R.~H.}\ \bibnamefont {Cohen}}, \ and\
  \bibinfo {author} {\bibfnamefont {A.}~\bibnamefont {Friedman}},\ }\href
  {http://stacks.iop.org/1749-4699/5/i=1/a=014019} {\bibfield  {journal}
  {\bibinfo  {journal} {Computational Science \& Discovery}\ }\textbf {\bibinfo
  {volume} {5}},\ \bibinfo {pages} {014019} (\bibinfo {year}
  {2012})}\BibitemShut {NoStop}%
\bibitem [{\citenamefont {Lehe}\ \emph {et~al.}(2016)\citenamefont {Lehe},
  \citenamefont {Kirchen}, \citenamefont {Andriyash}, \citenamefont {Godfrey},\
  and\ \citenamefont {Vay}}]{LeheCPC2016}%
  \BibitemOpen
  \bibfield  {author} {\bibinfo {author} {\bibfnamefont {R.}~\bibnamefont
  {Lehe}}, \bibinfo {author} {\bibfnamefont {M.}~\bibnamefont {Kirchen}},
  \bibinfo {author} {\bibfnamefont {I.~A.}\ \bibnamefont {Andriyash}}, \bibinfo
  {author} {\bibfnamefont {B.~B.}\ \bibnamefont {Godfrey}}, \ and\ \bibinfo
  {author} {\bibfnamefont {J.-L.}\ \bibnamefont {Vay}},\ }\href {\doibase
  http://dx.doi.org/10.1016/j.cpc.2016.02.007} {\bibfield  {journal} {\bibinfo
  {journal} {Computer Physics Communications}\ }\textbf {\bibinfo {volume}
  {203}},\ \bibinfo {pages} {66 } (\bibinfo {year} {2016})}\BibitemShut
  {NoStop}%
\bibitem [{\citenamefont {Andriyash}\ \emph {et~al.}(2016)\citenamefont
  {Andriyash}, \citenamefont {Lehe},\ and\ \citenamefont
  {Lifschitz}}]{AndriyashPoP2016}%
  \BibitemOpen
  \bibfield  {author} {\bibinfo {author} {\bibfnamefont {I.~A.}\ \bibnamefont
  {Andriyash}}, \bibinfo {author} {\bibfnamefont {R.}~\bibnamefont {Lehe}}, \
  and\ \bibinfo {author} {\bibfnamefont {A.}~\bibnamefont {Lifschitz}},\ }\href
  {\doibase http://dx.doi.org/10.1063/1.4943281} {\bibfield  {journal}
  {\bibinfo  {journal} {Physics of Plasmas}\ }\textbf {\bibinfo {volume}
  {23}},\ \bibinfo {eid} {033110} (\bibinfo {year} {2016}),\
  http://dx.doi.org/10.1063/1.4943281}\BibitemShut {NoStop}%
\bibitem [{\citenamefont {Birdsall}\ and\ \citenamefont
  {Langdon}(2004{\natexlab{b}})}]{Birdsall2004appE}%
  \BibitemOpen
  \bibfield  {author} {\bibinfo {author} {\bibfnamefont {C.}~\bibnamefont
  {Birdsall}}\ and\ \bibinfo {author} {\bibfnamefont {A.}~\bibnamefont
  {Langdon}},\ }\href {http://books.google.fr/books?id=S2lqgDTm6a4C} {\emph
  {\bibinfo {title} {Plasma Physics via Computer Simulation, Appendix E}}},\
  Series in Plasma Physics\ (\bibinfo  {publisher} {Taylor \& Francis},\
  \bibinfo {year} {2004})\BibitemShut {NoStop}%
\bibitem [{\citenamefont {Vay}\ and\ \citenamefont
  {Arefiev}(2014)}]{VayAAC2014}%
  \BibitemOpen
  \bibfield  {author} {\bibinfo {author} {\bibfnamefont {J.}~\bibnamefont
  {Vay}}\ and\ \bibinfo {author} {\bibfnamefont {A.}~\bibnamefont {Arefiev}},\
  }\href@noop {} {\bibfield  {journal} {\bibinfo  {journal} {AIP Conference
  Proceedings}\ }\textbf {\bibinfo {volume} {in press}} (\bibinfo {year}
  {2014})}\BibitemShut {NoStop}%
\bibitem [{\citenamefont {Vincenti}\ and\ \citenamefont
  {Vay}(2016)}]{Vincenti2015}%
  \BibitemOpen
  \bibfield  {author} {\bibinfo {author} {\bibfnamefont {H.}~\bibnamefont
  {Vincenti}}\ and\ \bibinfo {author} {\bibfnamefont {J.-L.}\ \bibnamefont
  {Vay}},\ }\href {\doibase http://dx.doi.org/10.1016/j.cpc.2015.11.009}
  {\bibfield  {journal} {\bibinfo  {journal} {Computer Physics Communications}\
  }\textbf {\bibinfo {volume} {200}},\ \bibinfo {pages} {147 } (\bibinfo {year}
  {2016})}\BibitemShut {NoStop}%
\bibitem [{\citenamefont {Huebl}\ \emph {et~al.}(2015)\citenamefont {Huebl},
  \citenamefont {Lehe}, \citenamefont {Vay}, \citenamefont {Grote},
  \citenamefont {Sbalzarini}, \citenamefont {Kuschel}, \citenamefont
  {Bussmann},\ and\ \citenamefont {Huebl}}]{openPMD}%
  \BibitemOpen
  \bibfield  {author} {\bibinfo {author} {\bibfnamefont {A.}~\bibnamefont
  {Huebl}}, \bibinfo {author} {\bibfnamefont {R.}~\bibnamefont {Lehe}},
  \bibinfo {author} {\bibfnamefont {J.-L.}\ \bibnamefont {Vay}}, \bibinfo
  {author} {\bibfnamefont {D.~P.}\ \bibnamefont {Grote}}, \bibinfo {author}
  {\bibfnamefont {I.}~\bibnamefont {Sbalzarini}}, \bibinfo {author}
  {\bibfnamefont {S.}~\bibnamefont {Kuschel}}, \bibinfo {author} {\bibfnamefont
  {M.}~\bibnamefont {Bussmann}}, \ and\ \bibinfo {author} {\bibfnamefont
  {A.}~\bibnamefont {Huebl}},\ }\href {\doibase 10.5281/zenodo.33624} {\enquote
  {\bibinfo {title} {openpmd 1.0.0: Initial release},}\ } (\bibinfo {year}
  {2015})\BibitemShut {NoStop}%
\end{thebibliography}%

\end{document}